\documentclass[apsrev4-1,aps,superscriptaddress,longbibliography,twocolumn]{revtex4-1}

\usepackage{colortbl}
\usepackage{color}
\usepackage{amsfonts,amsmath,amssymb}
\usepackage[dvipsnames]{xcolor}
\usepackage{tabularx,hhline,tikz,multirow,array}
\usepackage{bm,enumitem}
\usepackage{dcolumn,stmaryrd}
\usepackage[colorlinks=true,pdfborder={0 0 0},linkcolor=red,citecolor = blue]{hyperref}
\usepackage{comment}
\usepackage[normalem]{ulem}

\usepackage{xr}          
\def\beq{\begin{equation}}
\def\eeq{\end{equation}}
\def\bea{\begin{eqnarray}}
\def\eea{\end{eqnarray}}

\def\rv {{\bf r}}

\def\qv {{\bf q}}

\newcommand{\Sv}{\mathbf{S}}
\newcommand{\Dv}{\mathbf{D}}

\allowdisplaybreaks
\begin{document}
\title{Tuneable skyrmion and anti-skyrmion fluids via mechanical strain in chiral kagome lattice}

\author{Gonzalo dos Santos}
\affiliation{CONICET and Facultad de Ingenier\'ia, Universidad de Mendoza, 5500 Mendoza, Argentina}

\author{Flavia A. G\'omez Albarrac\'in}
\affiliation{Instituto de F\'isica de L\'iquidos y Sistemas Biol\'ogicos (IFLYSIB), UNLP-CONICET, Facultad de Ciencias Exactas, 1900 La Plata, Argentina}
\email{corresponding author: albarrac@fisica.unlp.edu.ar}
\affiliation{Departamento de Ciencias B\'asicas, Facultad de Ingenier\'ia, Universidad Nacional de La Plata, 1900 La Plata, Argentina}

\author{Ludovic D.C. Jaubert}
\affiliation{CNRS, Universit\'e de Bordeaux, LOMA, UMR 5798, 33400 Talence, France}

\author{Pierre Pujol}
\affiliation{Laboratoire de Physique Th\'eorique, Universit\'e de Toulouse, CNRS, UPS, France}

\author{Eduardo M. Bringa}
\affiliation{CONICET and Facultad de Ingenier\'ia, Universidad de Mendoza, 5500 Mendoza, Argentina}
\affiliation{Centro de Nanotecnolog\'ia Aplicada, Facultad de Ciencias, Universidad Mayor, 8580745 Santiago, Chile}

\author{H. Diego Rosales}
\affiliation{Instituto de F\'isica de L\'iquidos y Sistemas Biol\'ogicos (IFLYSIB), UNLP-CONICET, Facultad de Ciencias Exactas, 1900 La Plata, Argentina}
\affiliation{Departamento de Ciencias B\'asicas, Facultad de Ingenier\'ia, Universidad Nacional de La Plata, 1900 La Plata, Argentina}

\date{\today}

\begin{abstract}
Magnetic skyrmions are nanometric swirling spin textures that exhibit remarkable stability at finite temperatures, making them promising candidates for spintronic applications. Achieving controllable stability and transitions between distinct topological structures is crucial for practical implementations. In this work, we investigate the effect of uniaxial mechanical strain on a magnetic model on the kagome lattice, focusing on skyrmion stability and emergent topological phases. To this end, we consider a Heisenberg model that includes exchange interactions and both in-plane and out-of-plane Dzyaloshinskii-Moriya interactions. Using a combination of Spin-Lattice Dynamics and Monte Carlo simulations, we explore uniaxial strain variations in the range of $-10\%$ to $10\%$, showing important effects on the phase diagram. 
For compressive strain, we find that the density of skyrmions in the skyrmion gas (SkG) phase can be tuned and that the stability of this phase extends to higher temperatures. Tensile strain, in contrast, reduces the number of skyrmions and promotes transitions to other magnetic states. Within this regime, strain levels of about ($\sim4-6\%$) lead to a change in topological charge, turning skyrmions ($Q=-1$) into antiskyrmions ($Q=+1$).
We also examine how strain affects other phases commonly appearing in skyrmion-hosting systems, such as the helical and fully polarized states, showing that mechanical deformation alters their stability and characteristic properties.
Finally, we compare these results with the strain response of a more conventional skyrmion model, in order to clarify the role of the different interactions involved. Our results identify strain as an experimentally accessible route for engineering topological spin textures.
\end{abstract}

\maketitle
\section{Introduction}  
\label{sec:intro}

In the last decade, magnetic skyrmions -mesoscopic, spiraling structures with topological properties found in certain materials- have garnered significant attention due to their potential applications in next-generation storage devices and quantum computing \cite{raab2022brownian,zhang2015magnetic,pinna2018skyrmion,lee2023perspective}. It is well established that the formation of these topological textures typically arises from the interplay of competing interactions, such as exchange interactions and the Dzyaloshinskii-Moriya interaction (DMI) \cite{Moriya1960,dzyaloshinskii1964}, under an external magnetic field. These conditions are often present at the surfaces of magnetic layers. However, this is not the only mechanism. Skyrmions can also be stabilized through alternative pathways, including bond-dependent exchange anisotropy \cite{yi2009skyrmions,gao2020fractional,amoroso2020spontaneous,wang2021meron,rosales2022anisotropy}, the Ruderman-Kittel-Kasuya-Yosida (RKKY) interaction \cite{wang2020skyrmion},  dipolar interactions \cite{UtesovDipolar}, higher-order exchange interactions \cite{paul2020role} and magnetic frustration \cite{okubo2012multiple,mohylna2022spontaneous}.

The emergence and, most importantly, the stability of skyrmions are generally governed by a delicate balance among microscopic interactions for fixed temperature and magnetic field. Thus, tuning any of these interactions provides a means to control skyrmion phases, ranging from dense skyrmion lattices (SkL) to more dilute configurations, like skyrmion liquids and gases (Skyrmion Fluids (SkF)\cite{rosales2023skyrmion,gomez2024chiral}).  In the context of spintronic applications, it is useful to identify external parameters that allow one to adjust or stabilize skyrmion textures. Mechanical strain is one such parameter: several works have shown that uniaxial stress can create, suppress, or modify skyrmions in materials such as MnSi \cite{nii2015uniaxial}. Since this mechanism does not rely on electric or magnetic fields, it provides an alternative route for tuning magnetic textures in practical settings.

In addition, strain has a significant impact on the properties of skyrmions. In FeGe thin films, anisotropic strain induces distortions of individual skyrmions and modifies the skyrmion lattice \cite{shibata2015large,camosi2017anisotropic}. These effects originate from strain-driven changes in the DMI. Studies on Co/Pt multilayers further show that strain can tune the magnitude and anisotropy of the DMI \cite{deger2020strain,gusev2020manipulation,tanaka2020theoretical,feng2021field,zhang2021strain,el2022stability,li2022strain,littlehales2022enhanced,dong2023strain,mito2024magnetostriction}. In such systems, modifying the DMI affects the characteristic skyrmion size, their deformation, and their stability. Moreover, uniaxial strain has recently been shown to drive topological switching between skyrmions and bimerons in thin-film systems, underscoring the sensitivity of the topological charge to lattice deformation \cite{yang2024strain}.

Building upon these observations, we examine the effects of uniaxial strain on the stability and evolution of skyrmion phases in a magnetic model. We focus on a system in which the skyrmion density can be tuned with temperature\cite{rosales2023skyrmion,gomez2024chiral}, allowing skyrmions to disappear before paramagnetic fluctuations dominate. While strain may in principle influence both exchange and DM interactions, here we concentrate on its anisotropic effect on exchange couplings, which are directly sensitive to bond-length variations. Although previous studies have emphasized strain-induced modifications of the DMI via spin-orbit effects, our approach highlights a different mechanism: strain-induced anisotropy of the exchange interaction. The model also incorporates a competition between skyrmion-forming interactions and a classical chiral spin liquid background (CSL), known to stabilize unconventional skyrmionic phases \cite{rosales2023skyrmion,gomez2024chiral}. This enables us to examine strain-driven phenomena that do not appear in more conventional skyrmion models. For clarity, we later compare these results with the strain response of a system without the CSL contribution.

We then examine how uniaxial strain modifies the balance between exchange and DM interactions, and how this affects the transitions between dense skyrmion-lattice phases and more dilute skyrmion-gas regimes at finite temperature. This includes tracking how the phase boundaries shift under strain and how the stability range of the different skyrmion phases is modified.

The rest of the paper is structured as follows.
In Sec.~\ref{sec:metodos}, we introduce the magnetic model, discuss its strain-free behavior, and describe how strain is incorporated through the spin-lattice framework.
Sec.~\ref{sec:results} presents the strain-dependent analysis, including the evolution of skyrmion, antiskyrmion, and related phases, together with the comparison to the model without the CSL contribution.
Finally, Sec.~\ref{sec:conclusions} summarizes the main results and outlines possible extensions of this work.

\section{Model and Methods}
\label{sec:metodos}

\begin{figure}[thb]
\includegraphics[width=0.99\columnwidth]{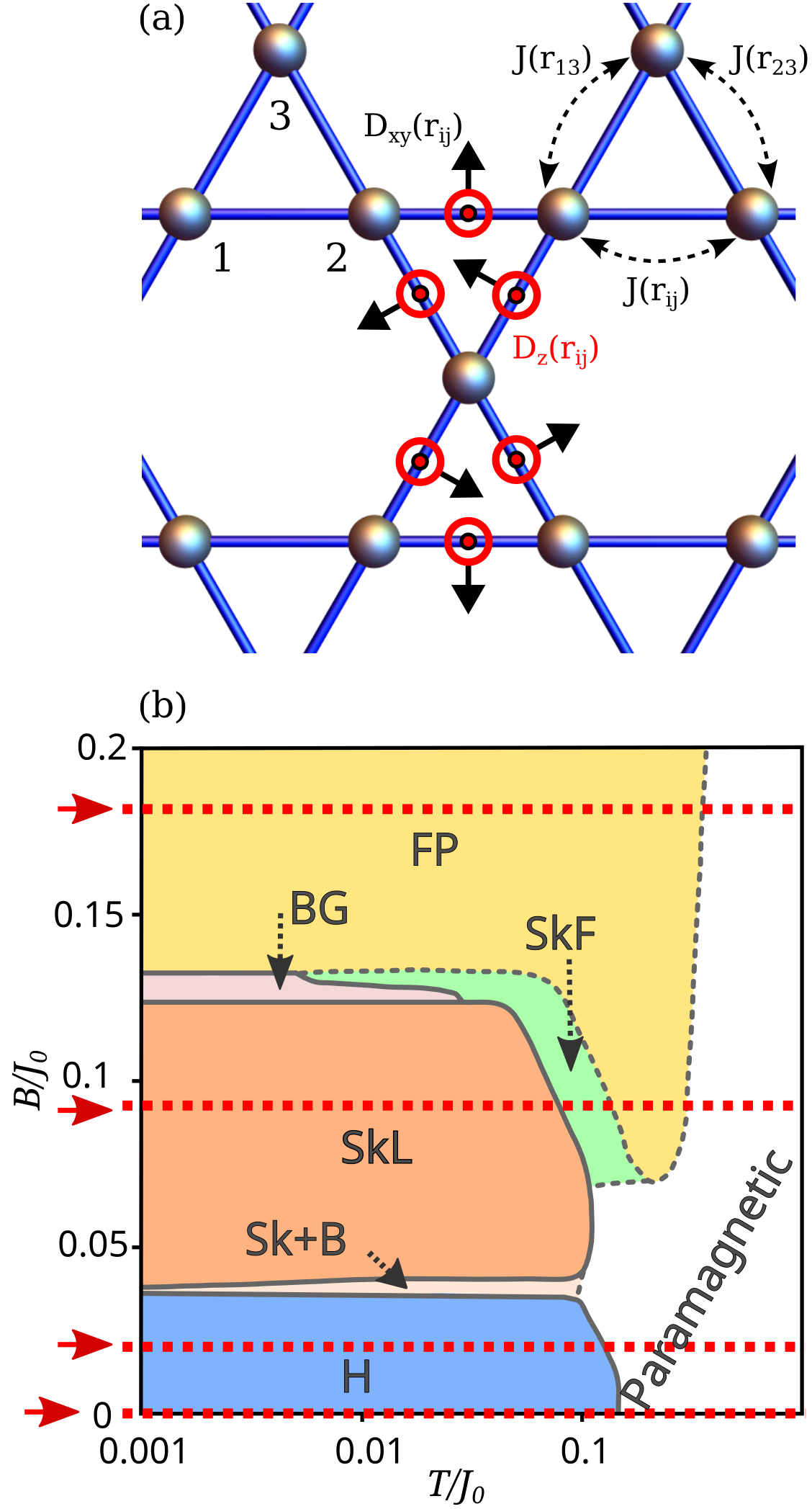}
\caption{(a) Kagome lattice and DM vectors. Labels $1,2,3$ represent the three sublattices. (b) Phase diagram in the $B-T$ plane  for $\epsilon=0$; red dashed lines marked with red arrows mark representative phases selected for further strain-dependent analysis.}
\label{fig:latt-J-PD}
\end{figure}

To investigate the effect of uniaxial strain, we use a dimensionless parameter $\epsilon$, taken as positive for compressive and negative for tensile deformations. In figures and captions, strain values are reported in percent, i.e. strain$(\%)=100\epsilon$. On the skyrmion phase, we consider $N$ atoms arranged in a kagome lattice (Fig.~\ref{fig:latt-J-PD}(a)), each possessing a classical magnetic moment $\Sv_i$. Inclusion of mechanical strain introduces modifications to the magnetic couplings, and by simulating the system under these conditions, we can directly assess how strain influences the stability and dynamics of skyrmion configurations.

%
\begin{figure*}[htb]
\includegraphics[width=0.99\textwidth]{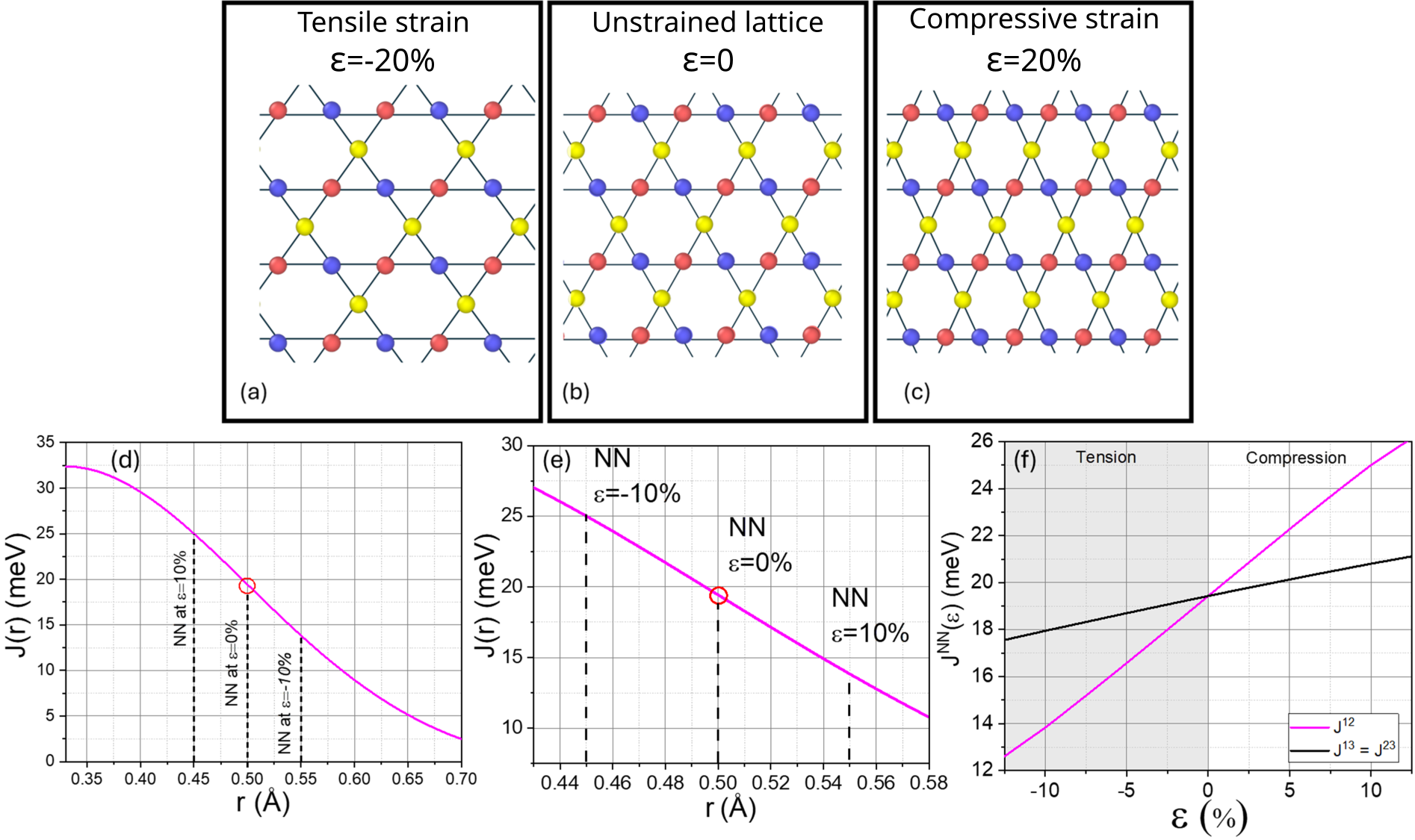}
\caption{(Color online) Representation of strain-induced anisotropy in the kagome lattice and its impact on magnetic exchange interactions. Top row: schematics of the deformed Kagome lattice. Atoms belonging to the sub-lattices 1, 2, 3 are represented as red, blue and yellow spheres respectively. The applied strain along the $x$ direction breaks the symmetry between sites (1, 2, 3), leading to anisotropic changes in bond lengths. In our simulations we have reached up to strains of $\pm  10\%$, nevertheless in (a) and (c) larger strains are included for visual purposes. Bottom row: panel (d) shows the fitted exchange function $J(r)$ (eq. \ref{eq: J(r)}). Vertical dashed-lines indicate the nearest-neighbor (NN) distance of the bond 1-2 for different degrees of deformation. In (e), a zoom of $J(r)$ in the strain range from $-10\%$ to $10\%$ is shown, evidencing a nearly linear behavior in that strain range. Panel (f) displays the strain dependence of NN exchange couplings $J^{12}(\epsilon)$ and $J^{13}(\epsilon)=J^{23}(\epsilon)$, revealing a stronger sensitivity for horizontal bonds compared to diagonal ones. This anisotropic response underpins the strain-tunable magnetic behavior discussed in the text.}
\label{fig:J_vs_strain}
\end{figure*}
\subsection{Magnetic Model (pure spin Hamiltonian)}
\label{sec:Model}
We begin by considering the magnetic part of the system, described by the Hamiltonian
\begin{eqnarray}
\mathcal{H}_{mag}&=&-\sum_{\langle ij\rangle}J(r_{ij})\;\Sv_i\cdot\Sv_j-\sum_{\langle ij\rangle}\Dv_{xy}(r_{ij})\cdot(\Sv_i\times\Sv_j)\nonumber\\
&&-\sum_{\langle ij\rangle}\Dv_{z}(r_{ij})\cdot(\Sv_i\times\Sv_j)-B\sum_i S^{z}_i,
\label{eq:HamSpin}
\end{eqnarray}
\noindent where $\mathbf{S}_i$ are classical Heisenberg spins of unit length ($|\mathbf{S}_i|=1$) at site $i$ on the kagome lattice (Fig.~\ref{fig:latt-J-PD}(a)) and the sum $\sum_{\langle ij\rangle}$ runs over all distinct nearest-neighbor pairs. The functions $J(r_{ij})$, $\Dv_{xy}(r_{ij})$,  and $\Dv_z(r_{ij})$ denote the exchange, in-plane DM, and out-of-plane DM couplings between nearest neighbors separated by a distance $r_{ij}=|\mathbf{r}_j-\mathbf{r}_i|$.  The in-plane DM vector, $\Dv_{xy}$, lies perpendicular to each bond, whereas $\Dv_z$ is oriented out of the plane, as shown in Fig.~\ref{fig:latt-J-PD}. The last term accounts for the Zeeman coupling to an external field applied along $z$. Notice that for $\Dv_z=0$ this model reduces to a very well known skyrmion Hamiltonian, where nearest neighbors ferromagnetic exchange interactions and in-plane DM are sufficient to stabilize skyrmions under an external magnetic field \cite{yu2010real}.

The exchange and DM couplings depend on the interatomic distance and are therefore modified 
by strain. Before addressing this strain-dependent case, we briefly recall the strain-free behavior of the model when all couplings are fixed, as in Refs.~\cite{rosales2023skyrmion,gomez2024chiral}.
For that case, Eq.~(\ref{eq:HamSpin}) reproduces the strain-free phase diagram shown in Fig.~\ref{fig:latt-J-PD}(b).  The moel stabilizes a chiral spin liquid background from which several magnetic textures emerge: helical (H), mixed skyrmion–bimeron (Sk+B), skyrmion lattice (SkL), skyrmion fluid (SkF), bimeron glass (BG), and the field-polarized (FP) state.  
In Fig.~\ref{fig:latt-J-PD}(b), the dashed lines indicate representative magnetic-field values used later in the strain analysis.
A central feature of this Hamiltonian is the CSL regime: its large entropy generates an extended intermediate region in which skyrmions appear with a temperature-dependent density, forming a skyrmion gas before being suppressed by paramagnetic fluctuations.
This behavior does not occur in more conventional chiral models without the CSL contribution (e.g., with $D_z = 0$), where no high-temperature SkF phase is present.

\subsection{Spin-Lattice Dynamics}
Having introduced the magnetic model, we now incorporate lattice degrees of freedom through spin-lattice dynamics (SLD). In this approach, implemented with the SPIN package of LAMMPS \cite{tranchida2018massively,thompson2022lammps}, atomic positions evolve via molecular dynamics, while spin vectors follow the stochastic Landau-Lifshitz equation.

The full spin-lattice Hamiltonian is
\begin{eqnarray}
\mathcal{H}&=&\sum_{i=1}^N\frac{|{\bf p}_i|^2}{2m}+\sum^{N}_{i,j;i\neq j}V(r_{ij})+\mathcal{H}_{mag},
\label{eq:HamTot}
\end{eqnarray}
where the first term corresponds to the kinetic energy of the $N$ atoms, with the $i$-th atom having momentum ${\bf p}_i$ and mass $m_i$. The second term, $V(r_{ij})$ describes the interaction between atoms separated by a distance $r_{ij}$. \\
The $\mathcal{H}_{mag}$ term in Eq.~(\ref{eq:HamTot}) corresponds to the magnetic Hamiltonian, defined in Eq.~(\ref{eq:HamSpin}).

In this more general setting, the couplings entering $\mathcal{H}_{mag}$ depend on the instantaneous atomic positions.  
In particular, the exchange interaction becomes an explicit function of distance, $J(r_{ij})$, which allows the magnetic sector to respond to lattice distortions such as uniaxial strain.  For our simulations, $J(r_{ij})$ is modeled through the Bethe-Slater form
\begin{eqnarray}
J(r_{ij})&=&4a\left(\frac{r_{ij}}{d}\right)^2\left[1-b\left(\frac{r_{ij}}{d}\right)^2\right]e^{-\left(\frac{r_{ij}}{d}\right)^2}\Theta\left(R_c-r_{ij}\right), \nonumber\\
\label{eq: J(r)}
\end{eqnarray}
\noindent where $\Theta\left(R_c-r_{ij}\right)$ is the Heaviside step function and $R_c$ is the cutoff. 
The parameters $a$, $b$, and $d$ were tuned so that $J(r_{ij})$ behaves similarly to the exchange function of Fe \cite{pajda2001ab}, as shown in Figure \ref{fig:J_vs_strain}(d). This ensures that, when the lattice undergoes compression or tension due to a strain $\epsilon$, the relative percentage change in $J(r_{ij})$ (compared to its nearest-neighbor value) matches the corresponding percentage change observed in the exchange function of Fe. 
We note that exchange interactions can exhibit non-monotonic behavior in specific materials due to complex electronic contributions.  Nevertheless, modeling $J(r_{ij})$ using a Bethe-Slater functional provides a robust effective description for 3d transition metals \cite{cardias2017bethe} and fulfills the requirement of an analytical and differentiable functional necessary for the self-consistent calculation of reciprocal forces within the SLD framework.

When uniaxial strain $\epsilon$ is applied along the $x$ axis, the three nearest–neighbor distances become inequivalent due to the deformation of the kagome triangle. From the lattice geometry (Fig.~\ref{fig:J_vs_strain}(a–c)), the strained bond lengths are: $r_{12}(\epsilon)=r_0(1-\epsilon)$ and $r_{13}(\epsilon)=r_{23}(\epsilon)=\frac{r_0}{2}\sqrt{\epsilon^2-2\epsilon+4}$. 
Inserting these distances into the Bethe–Slater form directly yields anisotropic exchange couplings: $J_{12}(\epsilon)$ varies more strongly with strain than $J_{13}(\epsilon)=J_{23}(\epsilon)$ (Fig.~\ref{fig:J_vs_strain}(f)). The explicit expressions for these couplings are provided in  Sec. I in the Supplemental Material (SM) \cite{SM}.

For the uniaxial deformation in this work, both DM interaction $\Dv_{xy}(r_{ij})$ and $\Dv_z(r_{ij})$ also depend on the distance between nearest neighbors, $r_{ij}$, because changes in that distance can modify the direction of the versor $\vec{e}_{ij}$ joining a spin pair $\Sv_i$,$\Sv_j$, and are therefore susceptible to strain-induced modifications. While mechanical strain can, in principle,  influence both magnetic interactions, in this work, we focus on systems where strain has a more pronounced effect on the exchange coupling, $J(r_{ij})$, than on the DM interaction. This assumption is motivated by the direct sensitivity of $J(r_{ij})$ to bond length variations, which typically dominate the response to strain. As shown in Figure S1 in the SM \cite{SM}, $\Dv(r_{ij})$ exhibits only minor variations along the bonds under strain. This reflects its weaker dependence on bond length changes, consistent with the predominant influence of spin-orbit coupling effects (see Sec. II from the SM). Such near-constancy of $\Dv(r_{ij})$ further justifies our focus on the anisotropic modifications in $J(r_{ij})$. 

While previous studies have emphasized the role of strain in modulating $\Dv(r_{ij})$ to influence spin textures via spin-orbit effects \cite{nii2015uniaxial,shibata2015large,shen2022strain,jiang2022tuning}, our approach is complementary. It focuses on the strain-induced anisotropy in $J(r_{ij})$ as the primary mechanism for stabilizing or destabilizing skyrmion phases. This perspective allows us to explore an alternative regime, complementing prior DM-focused analyses and broadening the understanding of how mechanical strain can be utilized to control skyrmionic states. A recent work \cite{zhang2025formation} has shown that exchange may exhibit larger strain tunability than interfacial DMI in certain in 2D chiral magnets, further motivating our focus on $J(r_{ij})$ as the primary strain-sensitive parameter in the present model. 

These anisotropic changes of the exchange couplings are expected to have two main consequences: first, the inequivalence between $J_{12}$ and $J_{13}=J_{23}$ introduces a preferred direction in the kagome lattice, which may manifest as anisotropic skyrmion textures under strain. Second, since strain alters the effective ratio $D_z/J$, compressive strain ($\epsilon>0$) shifts the system toward the regime $D_z/J<\sqrt{3}$, where, in the pure $(-J,D_z)$ model, ferromagnetic-like alignments are energetically favored \cite{gomez2024chiral}. Conversely, tensile strain ($\epsilon<0$) increases the effective ratio $D_z/J>\sqrt{3}$, favoring umbrella-like arrangements in that model \cite{gomez2024chiral}. The additional ingredient in our case is the presence of $D_{xy}$, which stabilizes skyrmions (or antiskyrmions), while these underlying tendencies set the stage for the strain-induced transitions discussed below.

In this framework, strain is assumed to primarily modify the exchange interaction through its dependence on interatomic distance, while the magnitude of the local magnetic moments is kept fixed. This approximation neglects possible magnetovolume and magnetoelastic effects, which are not included in the Hamiltonian and are therefore beyond the scope of this minimal description. Variations of the local magnetic moment under strain are expected to remain moderate compared to the stronger distance dependence of the exchange interaction \cite{paes2019strain}.

To complete the description of the SLD method used in our simulations, we note that the atomic and spin degrees of freedom evolve according to coupled Langevin equations. These equations govern the dynamics once the strain-modified couplings discussed above are specified. In the SLD framework, the atoms and spins evolution is governed by the following coupled Langevin equations \cite{tranchida2018massively},
\begin{equation}
\frac{d\bf{r}_i}{dt} = \frac{\bf{p}_i}{m_i} \label{EOM-p1}    
\end{equation}
\begin{equation}
   \frac{d\bf{p}_i}{dt} = \sum_{i,j,i\neq j}^{N} \left[-\frac{dV \left(r_{ij}\right)}{dr_{ij}} + \frac{dJ\left(r_{ij}\right)}{dr_{ij}} \bf{s}_i \cdot \bf{s}_j  \right]\bf{e}_{ij} - \frac{\gamma_L}{m_i} \bf{p}_i + \bf{\xi}_i  \label{eq:EOM-atoms} 
\end{equation}
\begin{equation}
    \frac{d\bf{s}_i}{dt} = \frac{1}{1+\lambda_s^2} \left[ \left( \bf{\omega_i} + \bf{\zeta}_i \right) \times \bf{s}_i + \lambda_s\bf{s}_i \times \left(\bf{\omega}_i \times \bf{s}_i\right)  \right]  \label{EOM-spins}
\end{equation}
Equation (\ref{eq:EOM-atoms}) describes the dynamics of the atoms, which are influenced by both spins orientation and the properties of the exchange function $ J(r_{ij})$. The parameter $\gamma_L$ represents the lattice damping coefficient, while $\xi(t)$ denotes a random fluctuating force sampled from a Gaussian distribution. The spin evolution is determined by Eq.~\ref{EOM-spins}, where the effective field acting on spin $i$ is expressed as $\omega_i = - \frac{1}{\hbar} \frac{\partial \mathcal{H}_{\text{mag}}}{\partial s_i}$. The parameter $\lambda_s$ corresponds to the Gilbert damping for spins, and $\zeta(t)$ represents a stochastic field following a Gaussian probability distribution. For further details on the Langevin functions and the stochastic field, please refer to \cite{dossantos2023hysteresis, tranchida2018massively}.

%
\subsubsection{Spin-Lattice Dynamics simulation parameters}

While Spin-lattice dynamics simulations offer a complete description of the dynamic coupling between the spins and the atomic lattice, in the present study, an approximation considering a ``frozen'' lattice was implemented. This means that, after the strain is applied, the condition that the atomic positions and velocities remain constant was imposed. This methodological choice is grounded in the experimental observation of magnetic skyrmions typically at low temperatures. At these temperatures, the available thermal energy is insufficient to significantly excite phonons, which implies that atomic vibrations have a minimal effect on the spin dynamics. This study corresponds to a situation of quasi-static deformation, relevant to many experiments. Future studies might include dynamic deformation, but due to atomistic simulation constraints because of sub-fs timesteps, the strain-rates would be large, above 10$^6$~s$^{-1}$.

In this work, we have simulated the $N=6912$ spins arranged in a $48 \times 48$ kagome lattice with periodic boundary conditions. The NN distance of the undeformed lattice is $r_0 = 0.5\,\text{\AA}$. The timestep adopted was 1 fs and the simulations spanned over 10$^6$ steps. These simulation times allows the magnetic energy and the magnetization of the system to reach steady values. The Gilbert damping in eq.\ref{EOM-spins} was set to $\lambda_s=0.01$.

All magnetic interactions and temperatures are expressed in units of $J_0$. Here, $J_0$ refers to the nearest-neighbor exchange interaction in the original, undeformed lattice, with a value of $J_0=19.43$ $meV$, following the aforementioned rescaling of the Fe-based model for the exchange. All findings are discussed relative to this exchange energy scale $J_0$ because this ensures that the phase transitions are robust features of the model's phase space and can be qualitatively mapped to different chiral magnets by adjusting the energy and length scales accordingly.
Within this framework, the parameters in our model are set as $D_{xy}/J_0 = 0.5$, and $D_z/J_0 = \sqrt{3}$. All interactions cutoff were set to 1st neighbors.
\begin{figure*}[t!]
\includegraphics[width=0.95\textwidth]{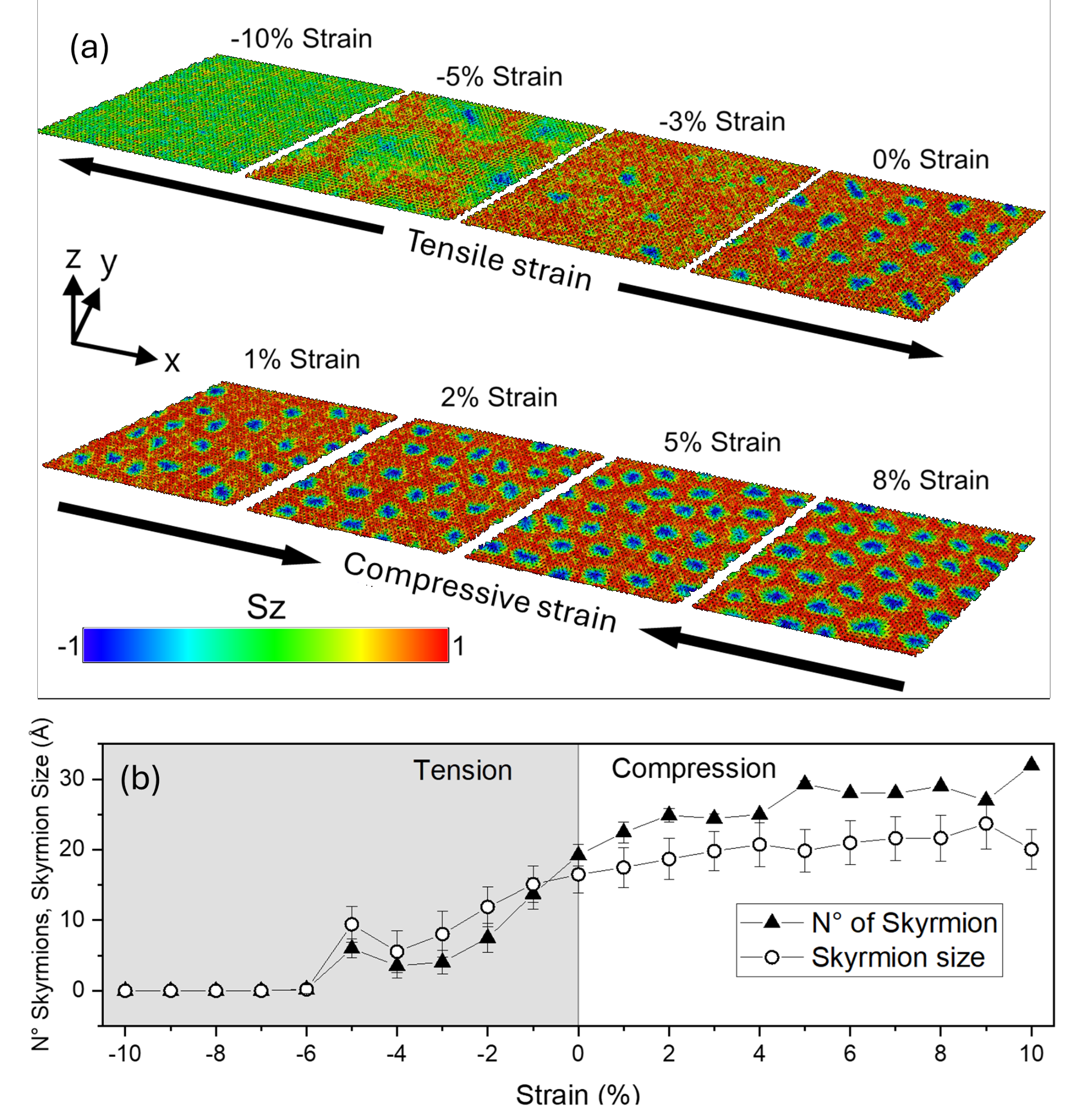}
\caption{(Color online) (a) Representative snapshots of the system at different degrees of strain. Individual spins vectors are colored according to their z component.  (b) Number of skyrmions and skyrmion size as a function of the applied strain.} 
\label{fig:Panel-sK-fluidPhase}
\end{figure*}
%

\subsection{Monte Carlo Metropolis}  
As a second approach, we performed Monte Carlo simulations using the Metropolis algorithm combined with microcanonical (over-relaxation) updates to improve sampling and equilibration. An annealing scheme was employed in which the temperature $T$ was gradually lowered at fixed external magnetic field $B$, allowing us to probe the equilibrium behavior across the different phases. Both $T$ and $B$ are expressed in units of the undeformed nearest-neighbor interaction $J_0$. 
Simulations were carried out on kagome lattices with $N = 3L^2$ sites, for sizes ranging from $L=12$ to  $L=108$, using periodic boundary conditions.  
Typical runs involved $10^5$--$10^6$ MC steps for equilibration and twice as many for measurements.

To incorporate strain, the exchange couplings were taken as $J_{ij}=J(r_{ij})$, where $r_{ij}$ are the strain-modified bond lengths.  
Unlike the Spin-Lattice Dynamics approach-where the lattice positions evolve self-consistently-here we use the pre-calibrated distance dependence $J(r)$ introduced earlier, details are given in Sec. II from the SM \cite{SM}.  
This allows strain to be included directly in the MC simulations while keeping the lattice static, providing complementary information on the thermal stability of the magnetic textures.

To characterize the phases, we computed the specific heat $C_v = (\langle E^2\rangle - \langle E\rangle^2)/T^2$, and two chirality observables:  
both are defined through the scalar triple product of three spins $\chi_{ijk} = \Sv_i\cdot(\Sv_j\times\Sv_k)$, but differ in the choice of plaquettes:  

(i) the nearest-neighbor chirality,
\[
\chi_{NN} = \frac{1}{8\pi N}\sum_{\triangle_{NN}}\Sv_i\cdot(\Sv_j\times\Sv_k),
\]
computed over all elementary kagome triangles;  

(ii) the sublattice chirality,
\[
\chi_Q = \frac{1}{4\pi (N/3)}\sum_{\triangle_{Q}}\Sv_i\cdot(\Sv_j\times\Sv_k),
\]
defined on the larger triangles belonging to a single kagome sublattice.  
The two quantities capture different and complementary aspects of the underlying topological order.

Finally, we also computed the static spin structure factor (the component perpendicular to the external magnetic field), defined as:
\[
S(\qv)=\frac{1}{N}\sum_{a=x,y} 
\Big\langle 
\big|\sum_j S^{a}_{j} e^{i\qv\cdot\rv_j}\big|^2
\Big\rangle.
\]
These quantities are useful because the CSL phase shows characteristic pinch-point patterns, while skyrmion lattices produce the familiar triple-$\qv$ structure.

\section{RESULTS AND DISCUSSION}
\label{sec:results}

We study the combined effects of strain, temperature, and magnetic field on the model defined in Eqs.~(\ref{eq:HamSpin}) and (\ref{eq:HamTot}). 
As a reference, in Fig.~\ref{fig:latt-J-PD}(b) we show the strain–free $B$–$T$ phase diagram obtained for $D_{xy}/J_0=0.5$ and $D_z/J_0=\sqrt{3}$, which includes the CSL background and the sequence of phases discussed earlier in section \ref{sec:Model}. 
Throughout this section, strain values are varied between $-10\%$ and $10\%$, corresponding to $-0.10 \leq \epsilon \leq 0.10$. Most of our analysis focuses on $B/J_0=0.09$, where both skyrmion lattice and skyrmion–fluid regimes are well represented, while several other field values hosting non–skyrmion phases are examined at the end of the section.

\subsection{Effect of Strain on Skyrmion-Lattice and Fluid Phases}
%
We start by considering the skyrmion–fluid state at $T/J_0=0.15$ and $B/J_0=0.09$, and examine how it responds to uniaxial strain along the $x$ axis. In Fig.~\ref{fig:Panel-sK-fluidPhase}(a) we show snapshots of representative equilibrated configurations under strain: from $\epsilon=0$ (SkF), compressive strain systematically increases both the number and size of skyrmions, whereas tensile strain reduces their density and size. In panel (b), these effects are quantified through the strain dependence of the number of skyrmions and the skyrmions' average size in the system.  
A singular peak appears near $\epsilon\!\sim\!-5\%$, associated with flattened magnetic textures that we later identify as antiskyrmions.  Although we have not performed a quantitative analysis of skyrmion diffusion, we qualitatively observe changes in their mobility under strain. See supplementary videos in the supplementary material.

Having established this behavior at one representative temperature, we now examine how strain affects skyrmion lattices and fluids across a wider temperature range.  
Figure~\ref{fig:snapsVsStrain-SkG} compiles real–space configurations and their corresponding structure factors $S(\mathbf{q})$ for several $(T,\epsilon)$ points in the skyrmion–hosting region of the diagram.

\subsubsection{Snapshots and Strain Dependence}

%
\begin{figure*}[t!]
\includegraphics[width=0.95\textwidth]{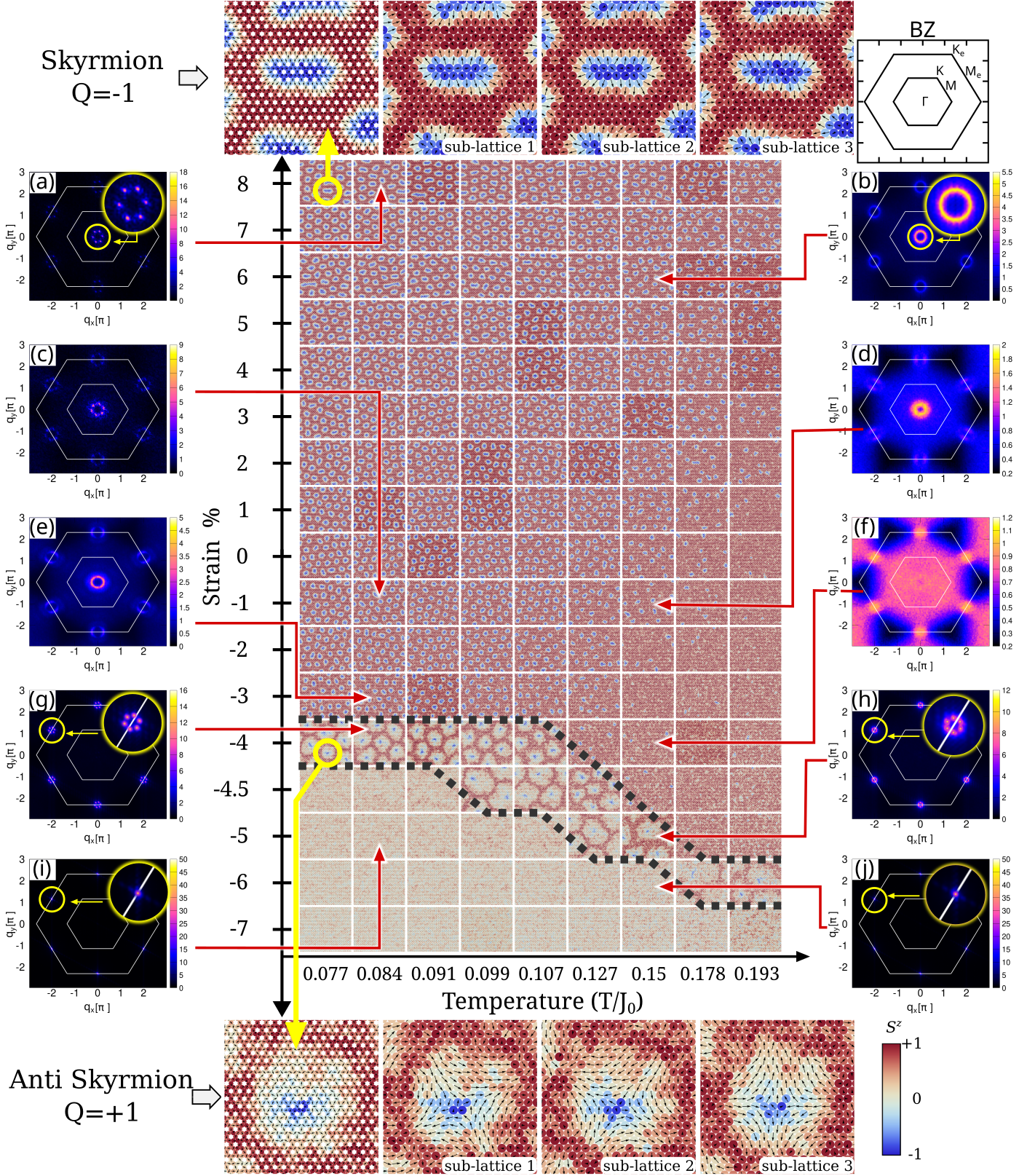}
\caption{(Color online) Strain-temperature phase diagram. The subpanels (a)-(j) surrounding the diagram correspond to the structure factors $S(\mathbf{q})$ in the reciprocal space $(q_x$, $q_y$ plane) for representative strain-temperature points. On the right side $(T/J_0 = 0.15$), as strain varies from $+8\%$ to $-7\%$, the system transitions through distinct configurations for the structure factor: a ring-like structure  corresponding to a skyrmion fluid, half-moon and pinch-point features characteristic of a skyrmion gas, a triple-$\mathbf{q}$ structure in the extended Brillouin zone (BZ) around the $M_e$ points associated with an antiskyrmion lattice, and finally, a single-$q$ umbrella-like structure at the $M_e$ points. On the left side ($T/J_0 = 0.084$), a similar sequence is observed as strain decreases from $+8\%$ to $-7\%$, but starting from a triple-$\mathbf{q}$ skyrmion lattice phase. These sub-panels illustrate the rich interplay between strain and temperature in shaping the magnetic textures and their reciprocal-space signatures.} 
\label{fig:snapsVsStrain-SkG}
\end{figure*}

The panels in the central part of Fig.~\ref{fig:snapsVsStrain-SkG} illustrate the evolution of the systems over a range of strain values from $-7\%$ to $+8\%$ and temperatures from $0.077 \leq T/J_0 \leq 0.193$.

\begin{itemize}

\item {\bf Compressive strain ($\epsilon > 0$):}\\

Compressive strain elongates skyrmions at the lowest temperature ($T/J_0=0.077$), leading to an anisotropic skyrmion lattice. This is evident in the topmost snapshots (labeled as ``Skyrmion $Q=-1$''), where each sublattice in the kagome structure hosts elongated skyrmions with a topological charge $Q=-1$. 
This elongation can be traced back to the anisotropic modification of the exchange couplings under uniaxial strain. In particular, strain along the $x$ direction enhances $J_{12}$ more strongly than $J_{13}$ and $J_{23}$. Since the characteristic wavelength of a single-$q$ spiral scales as $\lambda \sim J/D$, the mode ($q^*_x$) aligned  with the strain direction ($q_x$) acquires a longer wavelength. Similar strain-induced morphological changes have been reported in micromagnetic simulations of Fe$_3$GaTe$_2$ under uniaxial distortion \cite{zhang2025uniaxial}, supporting the role of strain as an efficient tuning parameter for chiral spin textures. Furthermore, recent MC  work \cite{zhang2025formation} has demonstrated that anisotropic exchange can generate elliptical skyrmions and that the exchange coupling can be more strongly tuned by strain than the DMI, reinforcing our argument that the elongation observed here originates primarily from strain-induced modifications of $J(r_{ij})$.

At higher temperatures ($T/J_0 > 0.15$), the effect of compressive strain manifests less through the geometry of individual skyrmions and more through their number. Strain enhances the effective exchange couplings, which can be understood as deepening the energy minima associated with skyrmion states. In this sense, compressive strain effectively counteracts thermal fluctuations, stabilizing skyrmions at temperatures where they would otherwise melt. This behavior is reflected in the increasing skyrmion density observed in the real space snapshots. This indicates that positive strain not only modifies the geometry of the SkL but also increases the number of skyrmions. 

\item {\bf Tensile strain ($\epsilon < 0$):}\\

Tensile strain, at small values, first reduces the number of skyrmions, 
an effect that becomes more pronounced as the temperature increases. For $\text{Strain\, }\lesssim -4\%$, the system undergoes a transition to an antiskyrmion phase (ASk), visible in the bottommost snapshots (labeled ``Anti Skyrmion $Q=+1$''), where individual antiskyrmions appear in the kagome sublattices.  Inspection of the topological charge density in this phase is shown in  Fig. S2 in the SM \cite{SM}. Beyond  $\text{Strain\, }\lesssim -5\%$, the ASk transitions to a single-$\mathbf{q}$ umbrella-like phase, reflecting a breakdown of the triple-$\mathbf{q}$ ordering. This occurs in the range of temperatures where the undeformed system is in the skyrmion fluid phase, emerging from the chiral spin liquid background.

%
%


This umbrella phase, described in detail in the SM, Fig. S3 \cite{SM} is characterized by plaquette-level order: each triangular unit of the kagome lattice hosts a repeating spin structure composed of three spins with similar projection along the external magnetic field. This coherent alignment produces a texture reminiscent of an “umbrella”. This texture arises from the superposition of the three triangular sub-lattices that compose the kagome lattice, displayed individually in  Figs. S3 (b-d). Each triangular sublattice shows a distinct spin ordering with a well-defined in-plane orientation pattern. When combined, these three sublattices produces the characteristic plaquette-level ordering of the umbrella-like phase.

This sequence can be understood by recalling how strain modifies the effective ratio $D_z/J$. For compressive strain ($\epsilon>0$), the system is pushed into the $D_z/J < \sqrt{3}$ regime, where ferromagnetic-like backgrounds are favored, and the presence of $D_{xy}$ then stabilizes skyrmions. In contrast, tensile strain ($\epsilon<0$) drives the system into the $D_z/J > \sqrt{3}$ regime, where umbrella-like order is preferred; under these conditions, the additional $D_{xy}$ interaction promotes the formation of antiskyrmions as a compromise texture. For sufficiently strong tensile strain, the dominance of $D_z$ over $J$ ultimately stabilizes the umbrella-like state.

In this sense, strain effectively tunes the balance between $D_z$ and $J$, allowing control over the transition from skyrmions to antiskyrmions, with the umbrella phase as the asymptotic limit at large negative strain.
\end{itemize}

\subsubsection{Structure Factor Evolution}
The structure factor $S(\mathbf{q})$ panels in Fig.~\ref{fig:snapsVsStrain-SkG} provide a complementary view of the ordering under strain. 

\begin{itemize}

\item {\bf Compressive strain ($\epsilon > 0$):}\\
At low temperature ($T/J_0=0.077$), for strain $=+8\%$, the SkL phase retains the triple-$\mathbf{q}$ signature around the $\Gamma$ point of the BZ, but the three wavevectors are no longer equivalent. The component aligned with the strain axis shifts to smaller $|\mathbf{q}|$, consistent with a longer spiral wavelength $\lambda \sim J/D$ in that direction. This imbalance directly explains the elongation of skyrmions seen in real space: the anisotropy of the textures is mirrored by the distortion of the reciprocal-space pattern. At higher temperatures ($T/J_0>0.1$), with increasing positive strain, these evolve into ring-like patterns, marking the crossover to a dense skyrmion liquid.

\item {\bf Tensile strain ($\epsilon < 0$):}\\
As strain decreases, the SkL recovers a more symmetric triple-$\mathbf{q}$ structure near $\epsilon \sim 0$. For $\epsilon \lesssim -4\%$, $S(\mathbf{q})$ emerges a triple-$\mathbf{q}$ around every high-symmetry  $M_e$ points, characteristic of the ASk phase. Beyond $\epsilon \lesssim -5\%$, this collapses into a single-$\mathbf{q}$ signal, indicative of the umbrella-like phase. This evolution of the spin textures is consistent with the strain-driven increase of the effective $D_z/J$, which pushes the system toward umbrella-type order, with antiskyrmions appearing as an intermediate configuration stabilized by the in-plane DM interaction.

At higher temperatures ($T/J_0>0.1$), the sequence changes: starting from $\epsilon=0$, $S(\mathbf{q})$ shows pinch-point features typical of diluted skyrmion fluids \cite{rosales2023skyrmion,gomez2024chiral}. For negative strain, the system reorganizes into triple-$\mathbf{q}$ peaks at the $M_e$ points, signaling the ASk phase. All this sequence of structure factor patterns shows how strain shifts the balance between exchange and DM couplings, driving the system between ferro-like  (skyrmions) and umbrella-like (antiskyrmions) regimes.

\end{itemize}

In short, strain reshapes both the geometry and topology of the textures. A more quantitative picture of these changes is obtained from the chirality analysis, discussed next.

\begin{figure*}[t!]
\includegraphics[width=0.9\textwidth]{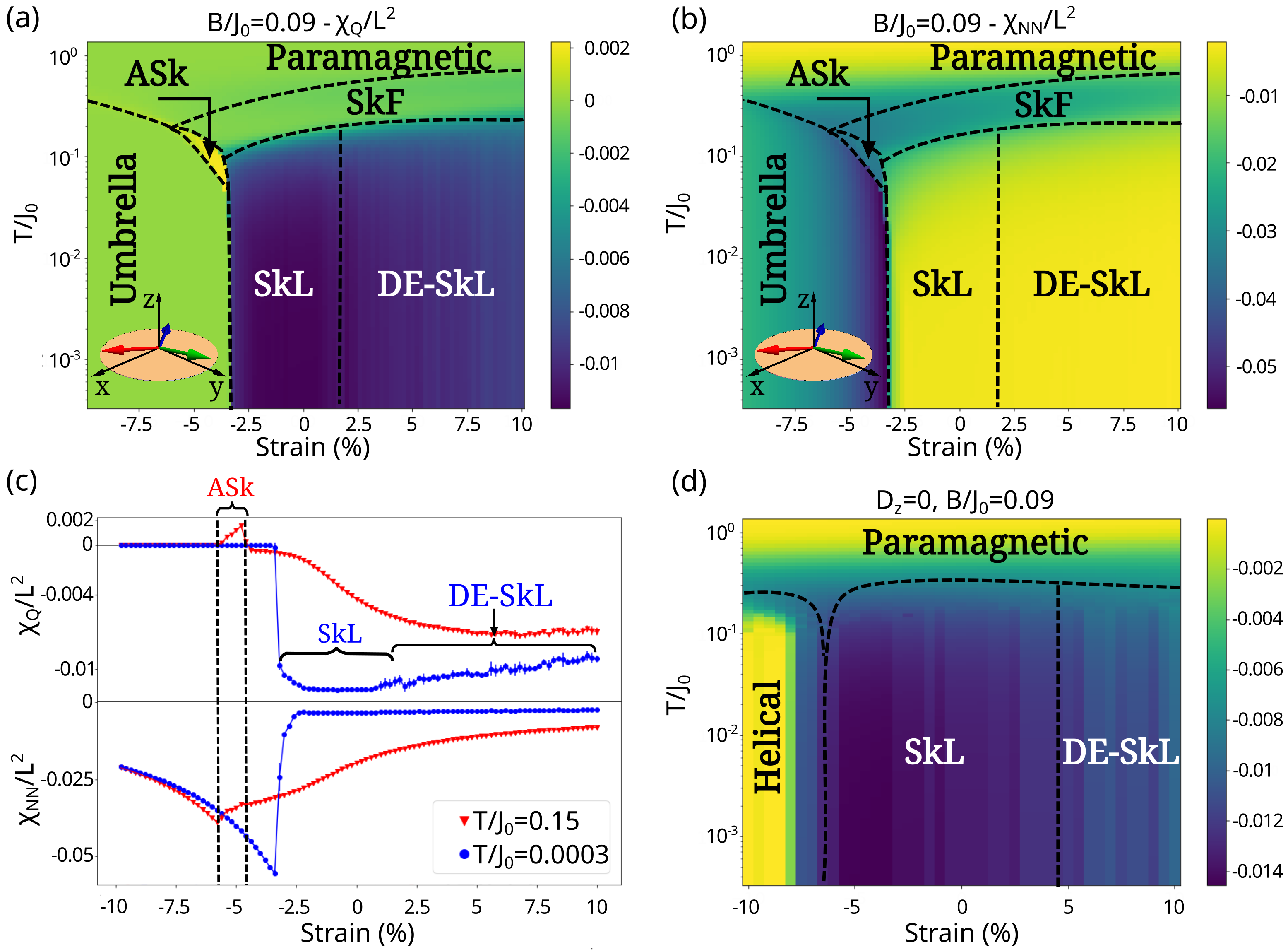}
\caption{(Color online) 
Chirality analysis as a function of strain and temperature at $B/J_0=0.09$. 
(a) Density plot of the sublattice chirality $\chi_Q/L^2$, indicating the different phases: Diluted Elongated Skyrmion Lattice (DE-SkL), Skyrmion Lattice (SkL), Skyrmion Fluid (SkF), Antiskyrmion (ASk), and Umbrella-like phase. The sign change of $\chi_Q$ signals the transition from SkL to ASk under tensile (negative) strain. 
(b) Density plot of the nearest-neighbor chirality $\chi_{NN}/L^2$ for the same parameters, which highlights the onset of the umbrella-like $q=0$ phase at large negative strain. 
(c) Line cuts of $\chi_Q/L^2$ (top) and $\chi_{NN}/L^2$ (bottom) as a function of strain for $T/J_0=0.15$ and $T/J_0=0.0003$, illustrating how positive strain reduces $\chi_Q$ (DE-SkL) while negative strain drives the system into ASk and umbrella-like states. 
(d) Density plot of $\chi_Q/L^2$ for the case $D_z=0$, showing the absence of SkF, ASk, and umbrella-like phases, leaving only SkL and helical states.}
\label{fig:SkG2strain}
\end{figure*}
\subsubsection{Chirality Phase Diagram: Insights into Strain-Induced Topological Transitions}
To move beyond the descriptive analysis of snapshots and reciprocal-space patterns, we now turn to a more global characterization of the system based on chirality. We construct a $T$ vs. strain phase diagram, computing both the sublattice chirality $\chi_Q$ and the nearest-neighbor chirality $\chi_{NN}$, which together allow us to distinguish skyrmionic, antiskyrmionic, and umbrella-like states. As discussed in previous works \cite{rosales2023skyrmion,gomez2024chiral}, $\chi_Q$ is a more suitable order parameter than the usual first-neighbor chirality $\chi_{NN}$ to capture emergent skyrmionic textures in this model with a chiral spin liquid background. The results for $B/J_0=0.09$ are summarized in Fig.~\ref{fig:SkG2strain} and show the following features:  
\begin{itemize}
\item {\it Diluted Elongated Skyrmion Lattice} (DE-SkL):\\
This phase occurs at positive values of strain $\epsilon$, particularly at low temperatures. Here, the skyrmions are not only elongated due to strain-induced anisotropy but also exhibit a more diluted nature, with a decreased density. This dilution of skyrmions is a direct consequence of the positive strain, as the lattice becomes increasingly anisotropic and the skyrmion cores stretch, see snapshots at top panel in Fig.~\ref{fig:snapsVsStrain-SkG}. 
\item {\it Skyrmion Lattice} (SkL):\\
This phase is characterized by a stable and ordered skyrmion structure. It is found in a broad region of the phase space, particularly at intermediate temperatures and strain values. In this region, the skyrmions form a well-defined periodic structure, and the chirality remains constant, indicating a uniform topological arrangement.
\item {\it Skyrmion Fluid} (SkF): \\
At higher temperatures and positive strain, the skyrmion lattice transitions into a disordered, fluid-like state. The SkF phase is associated with increased temperature and strain, where the skyrmions become more mobile and lose their ordered arrangement. This phase exhibits a decrease in the topological order, as seen in the gradual reduction of $\chi_Q$.
\item {\it Antiskyrmion Lattice} (ASk): \\
As strain becomes negative ($\epsilon < 0$), the system transitions to an antiskyrmion lattice. This phase is marked by a sign change in the sublattice chirality $\chi_Q$, which switches from negative to positive values. The ASk phase represents a topologically non-equivalent state, where the skyrmions are replaced by antiskyrmions, not as a simple substitution but as a consequence of the strain-driven shift in the effective $D_z/J$ ratio, moving the system from local ferro-like toward umbrella-like tendencies. In this context, snapshots confirm that antiskyrmions are more extended than skyrmions at higher $T$, with cores surrounded by umbrella-like plaquettes showing larger in-plane components, see snapshots in Figs.~\ref{fig:Panel-sK-fluidPhase} and S2.

\item {\it Umbrella-like Phase}:\\ 
At even more negative strain values ($\epsilon \ll 0$), the system enters the umbrella-like phase, characterized by a single-$q$ ordering. In this case, each triangular sublattice is formed by spins canted in the same direction, leading to $\chi_Q \approx 0$. However, as shown in Fig.~\ref{fig:SkG2strain}(c), the nearest-neighbor chirality $\chi_{NN}$ remains finite. The emergence of this phase is expected for sufficiently strong tensile strain, since reducing the exchange coupling effectively increases the ratio $D_z/J$. The resulting $q=0$ order is closely related to the planar $q=0$ configuration found at zero field for large $D_z$ \cite{essafi2017}; under finite field, the spins cant to form the umbrella phase with $\chi_{NN}\neq 0$.
\end{itemize}

The chirality $\chi_{NN}$ in Fig.~\ref{fig:SkG2strain}(b) supports the previous. Coming from the paramagnetic phase, $|\chi_{NN}|>0$ in the CSL regime. At lower temperatures, $\chi_{NN}$ remains close to zero in the SkL and DE-SkL phases that are stabilized with positive strain. However, for strong enough tensile strain (strain $\lesssim -2.5\%$), $|\chi_{NN}|$ increases sharply before decreasing smoothly as strain is further enhanced. This behavior reflects the progressive stabilization of the umbrella-like $q=0$ phase, which becomes less magnetized as $D_z/J$ grows.

In Fig.~\ref{fig:SkG2strain}(c), we further explore the strain-dependent behavior of $\chi_Q$ for two fixed temperature values, an intermediate one $T/J_0=0.15$ and the lowest simulated temperature $T/J_0=0.0003$. In a general overview, on the one hand, starting from positive strains, we see that $\chi_Q$ decreases progressively, with the SkL phase transitioning into the DE-SkL, reflecting a change in the skyrmion density and elongation. On the other hand, in the most negative strain regime, the system enters the $q=0$ phase, where the sublattice chirality $\chi_Q$ approaches zero. For a narrow but finite range of temperatures, between these two regimes $\chi_Q$ shows an abrupt change in sign, marking the emergence of the ASk phase, as seen for the $T/J_0=0.15$ curve. This phase is stabilized in temperatures where, in the strain-free model, skyrmions start to emerge from the chiral spin liquid background (SkF phase). The snapshots in Fig.~\ref{fig:Panel-sK-fluidPhase} shows that these antiskyrmions structures are more extended than the skyrmions found at higher temperatures, and in fact the core is surrounded by a significant amount of the umbrella-like plaquettes, characterized by a lower magnetisation and larger in-plane components than the field-polarized chiral spin liquid background, see Fig. S2 in the Supplementary Material \cite{SM}. This suggests that the ASk arises as a result of a delicate balance in the competition between a stronger tensile strain, the chiral spin liquid background and the in-plane DM interaction. The $\chi_{NN}$ curves as a function of strain for the same temperatures in Fig.~\ref{fig:SkG2strain}(c) are consistent with this analysis, although there is no clear indicator of the ASk phase, except for a possible change of curvature in the $\chi_{NN}/L^2$ curve at strain $\sim -4.5\%$.  These results as a function of the effective parametrisation of the exchange interaction are presented in  Fig. S4 from the SM \cite{SM}.

\begin{figure}[h!]
\includegraphics[width=1.0\columnwidth]{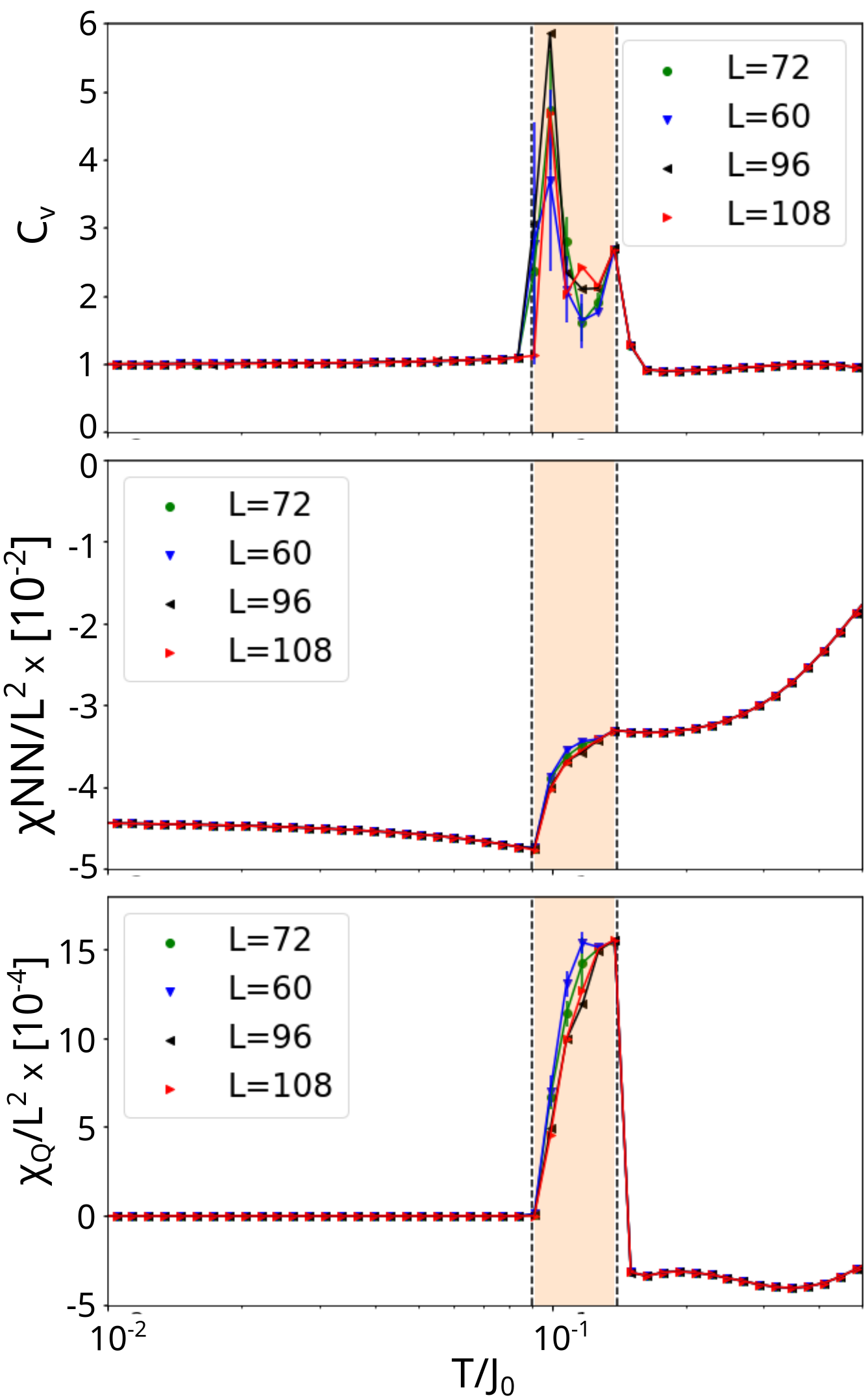}
\caption{Specific heat (top), nearest neighbor chirality $\chi_{NN}/L^2$ (middle) and sublattice chirality density $\chi_Q/L^2$ (bottom)  as a function of temperature for tensile strain $= -4.5\%$ ($\epsilon=-0.045$), for different lattice sizes $L=60,72, 96, 108$, and magnetic field $B/J_0=0.09$. Curves are obtained averaging over 5 independent realizations. When not seen, error bars are the size of the markers. The colored region corresponds to the ASk phase.}
\label{fig:StrainM45Ls}
\end{figure}

To gain a deeper insight in the formation and robustness of the ASk phase, we fix the strain to $\epsilon=-4.5\%$, and in Fig.~\ref{fig:StrainM45Ls} we present the behaviour of the specific heat,  nearest neighbor chirality $\chi_{NN}/L^2$ and sublattice chirality density $\chi_Q/L^2$ as a function of temperature, zooming in the temperature range $0.01 \leq T/J_0 \leq 1$, for $5$ independent copies and two larger system sizes $L=60,72$.  

The specific heat shows that indeed there seems to be a two-step evolution. Coming from the chiral spin liquid at high temperatures, characterized by a dip in $C_v$, there is a first peak that signals the crossover from the chiral spin liquid into the skyrmion fluid. In this case, the system then rapidly enters the ASk phase, characterized by the second larger peak in $C_v$ and by an abrupt jump and change of sign of the sublattice chirality $\chi_Q$. This chirality vanishes as the temperature is lowered and the system settles in the umbrella-like $q=0$ phase, which in turn has a non-zero nearest neighbor chirality, as can be seen in the middle and bottom panels of  Fig.~\ref{fig:StrainM45Ls}.  Although we defer a more detailed study on the nature of the transitions, these curves support the idea that the competition between the chiral spin liquid background and the $q=0$ umbrella ordering favored by an effectively larger $D_z$ is crucial for the realisation of antiskyrmions at intermediate temperatures.

\subsubsection{Comparison with the $D_z = 0$ Case: Insights from the Phase Diagram}

Finally, to clearly contrast the influence of the out-of-plane ($D_z$) DM interaction, we present a sublattice chirality density $\chi_Q/L^2$ phase diagram for $D_z = 0$, shown in Fig.~\ref{fig:SkG2strain} (d). This comparison highlights several important differences:

\begin{itemize}
\item Absence of the Skyrmion Fluid (SkF): In the absence of the $D_z$ term, the Skyrmion Fluid phase does not appear. This suggests that the DM interaction is essential for stabilizing the SkF phase, which is otherwise absent when $D_z = 0$.
\item No Antiskyrmion Phase (ASk): Similarly, the ASk phase disappears without $D_z$, indicating that the inclusion of this term is crucial for supporting the formation of antiskyrmions under negative strain. The system without $D_z$ does not exhibit the transition from the SkL to the ASk phase, which further emphasizes the role of the DM interaction in enabling this topological transition.
\item No Umbrella-like Phase: The umbrella-like phase, which is a feature of the system under large negative strains, as expected, vanishes when $D_z = 0$. 
\end{itemize}
In summary, the comparison between the phase diagrams for $D_z = 0$ and $D_z \neq 0$ clearly demonstrates the significant impact of the DM interaction on the topology and stability of the skyrmion phases, particularly with respect to the Skyrmion Fluid, Antiskyrmion Lattice, and umbrella-like phases. 

\subsection{Non-Skyrmion Phases}

Moving forward, we apply similar strain variations to other phases depicted in Fig.~\ref{fig:latt-J-PD}(b) at different magnetic fields, where no skyrmions are stabilized at lower temperatures: the helical (H) and the field polarized (FP) phases. We aim to assess whether the strain-induced modifications observed in the SkG phase at intermediate temperatures -such as the emergence of antiskyrmions under tensile conditions- are unique to this phase or represent a broader phenomenon across the phase diagram.

Inspecting the phase diagram presented in Fig.~\ref{fig:latt-J-PD}, we take $B/J_0=0.18$ for the field polarized phase, to ensure that the system is away from the skyrmion-like phases and deep into this FP phase. In this particular model, the FP phase is a non trivial one, since it stems from a chiral spin liquid. In order to study the effects of strain in the helical region, we focus on $B/J_0=0$ and $B/J_0=0.02$, where the system is closer to the chiral spin liquid higher temperature region, and under a magnetic field broken helices and bimerons are formed in a small region of intermediate temperatures, before the system sets in the topologically trivial helical phase. 

\begin{figure}[h!]
\includegraphics[width=1.0\columnwidth]{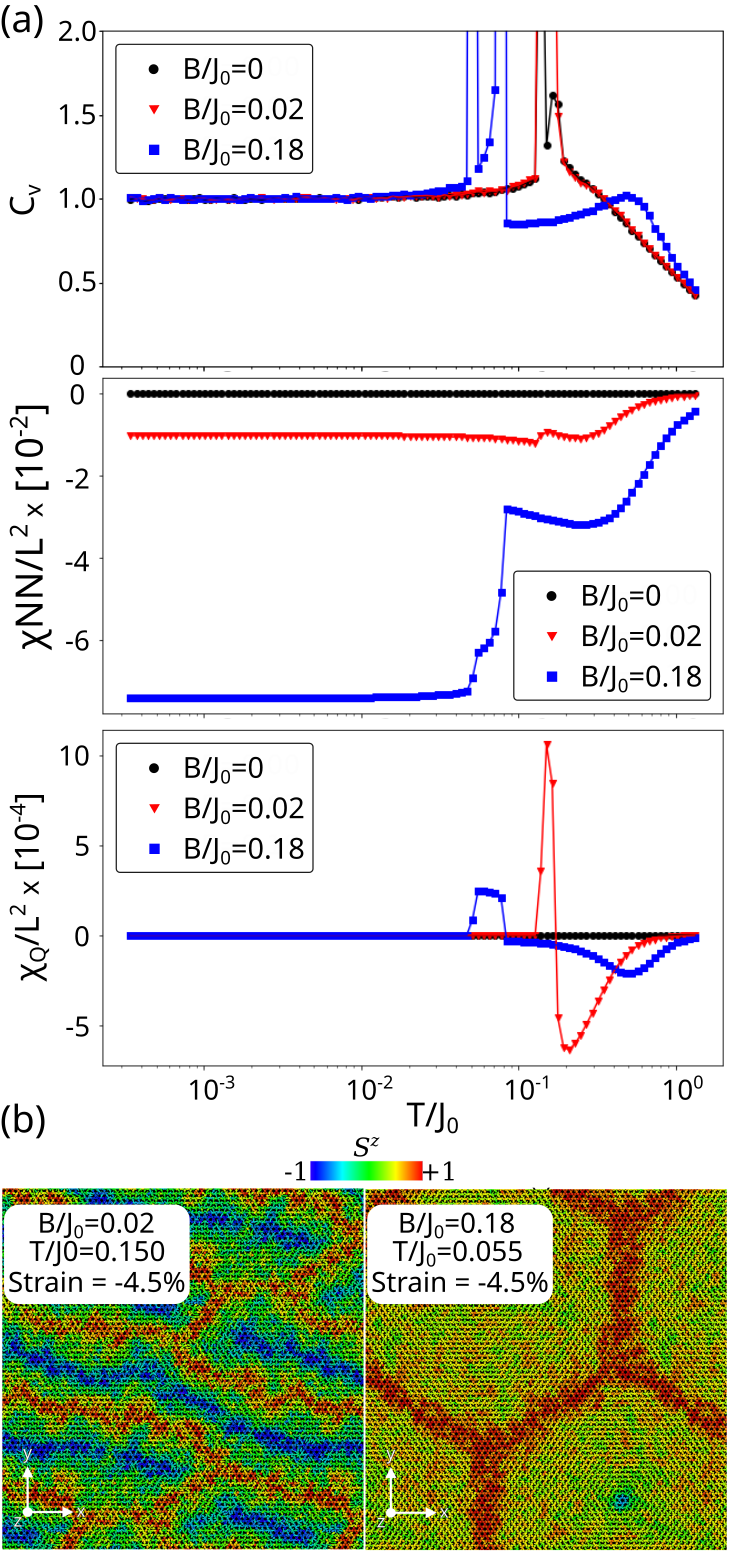}
\caption{(a)): Specific heat, nearest neighbor chirality $\chi_{NN}/L^2$  and sublattice chirality density $\chi_Q/L^2$ as a function of temperature for tensile strain $= -4.5\%$, and different values of the external magnetic field  $B/J_0=0,0.02,0.18$. (b): real space details of the different ``antiskyrmion'' phases at intermediate temperatures, for $T/J_0=0.15, 0.16$, for $B/J_0=0.02$ and $T/J_0=0.055, 0.077$, for $B/J_0=0.18$ }
\label{fig:StrainM45Bs}
\end{figure}

Following the discussion in the previous section, if antiskyrmions arise, they will do so at intermediate temperatures and for significant negative strain. Thus, we fix strain $=-4.5\%$ ($\epsilon=-0.045$) and study in Fig.~\ref{fig:StrainM45Bs}(a) the behavior with temperature of the specific heat, the sublattice chirality density and nearest neighbor chirality density for these three values of the external magnetic field $B/J_0=0.,0.02, 0.18$.  The specific heat curves suggest sharp transitions, and the chiral spin liquid behaviour is noticeable for $B/J_0=0.18$. For $B/J_0=0.02$, although there is no dip in the specific heat signaling the onset of the chiral spin liquid, the $\chi_{NN}/L^2$ curve show that the nearest neighbor chirality, associated with the chiral spin liquid background, is non-zero before the sharp peak in the $C_v$, whereas  $\chi_{NN}/L^2$  remains zero in all the temperature range for $B=0$. As in Fig.~\ref{fig:StrainM45Ls}, the region in temperature corresponding to the sharp peak in the $C_v$ for $B/J_0=0.02$ and $B/J_0=0.18$ matches the temperature range where $\chi_Q/L^2>0$, compatible with the stabilization of antiskyrmions. This spike in $\chi_Q/L^2$ is not present at zero field, further supporting the claim that the antiskyrmions emerge as a competition between the $q=0$ umbrella ordering, favored by a relatively larger $D_z$, and the chiral spin liquid found at higher temperatures. 
Finally, we examine the real-space configurations associated with the $\chi_Q>0$ region, shown in the bottom panels of Fig.~\ref{fig:StrainM45Bs}(b). 
For $B/J_0=0.02$, close to the helical phase boundary, the antiskyrmion-like textures are more elongated and less well defined, resembling the bimeron-like objects observed at higher temperatures. 
In contrast, for $B/J_0=0.18$ the system hosts bigger and more clearly defined antiskyrmions.

\begin{figure*}[htb!]
\includegraphics[width=1.0\textwidth]{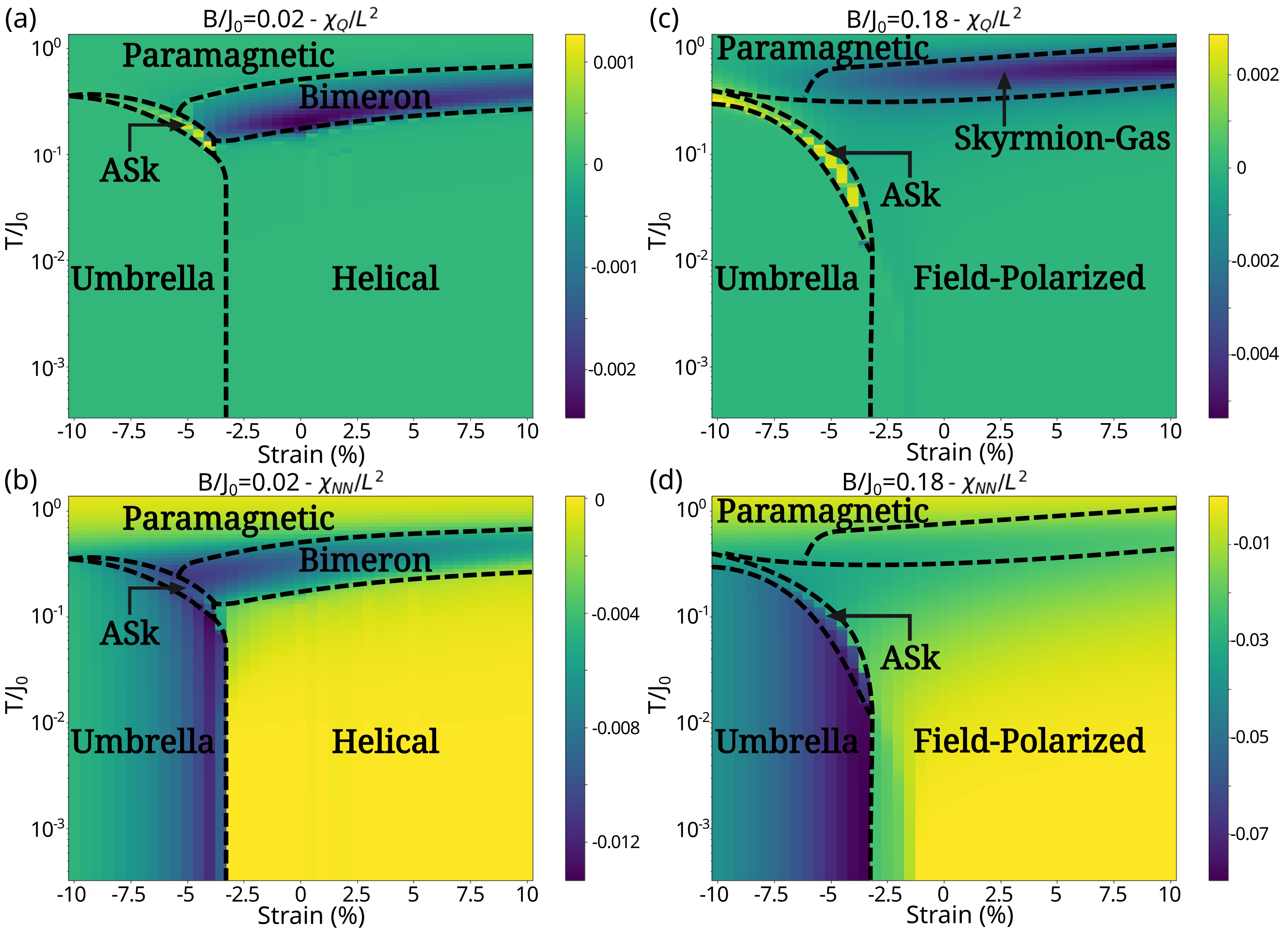}
\caption{Sublattice chirality density $\chi_Q/L^2$ (top) and nearest neighbor chirality density $\chi_{NN}/L^2$ (bottom) as a function of temperature and strain for external magnetic field  $B/J_0=0.02$ (left) and $B/J_0=0.18$ (right)}
\label{fig:StrainB002B018}
\end{figure*}

In Fig.~\ref{fig:StrainB002B018} we show density plots of the sublattice and nearest–neighbor chiralities for $B/J_0=0.02$ and $B/J_0=0.18$. 
For both fields, $\chi_Q/L^2$ remains essentially zero at low temperatures, indicating the absence of strain–induced skyrmion-like textures in this regime. 
At higher temperatures and under tensile strain, a sign change in $\chi_Q/L^2$ appears, although the region where this occurs is noticeably smaller than for $B/J_0=0.09$. 
This agrees with the real-space analysis: for these fields, the antiskyrmion-like configurations are less well defined and occupy a narrower parameter range. 
For sufficiently negative strain, both fields eventually enter the $q=0$ umbrella regime, reflected in the finite low-temperature values of $\chi_{NN}/L^2$ seen in the lower panels of Fig.~\ref{fig:StrainB002B018}.

\section{Summary and conclusions}
\label{sec:conclusions}

In this work, we have examined the effects of mechanical strain on the magnetic properties of skyrmion phases within a chiral Heisenberg model on the kagome lattice. By combining spin-lattice dynamics and Monte Carlo simulations, we have shown how strain influences the stability, topology, and dynamics of skyrmion phases, providing valuable insights into possible mechanisms for their control and manipulation in spintronic applications.

Our findings for the skyrmion lattice and fluid phases show that strain acts as a powerful tuning parameter. At low temperatures, compressive strain elongates skyrmions, introducing significant anisotropy into the skyrmion lattice. Conversely, tensile strain drives a transition from a skyrmion fluid to antiskyrmion lattices, altering the chirality of the system. Furthermore, strain extends the temperature range over which skyrmion-like configurations remain stable, allowing their persistence at conditions where they would otherwise disappear in the unstrained system. These results demonstrate that
strain not only modifies the symmetry and topology of skyrmionic textures but also enhances their thermal robustness.

In addition to modifying stability, strain also induces topological transitions within the skyrmion phases. Compressive strain can increase skyrmion density, facilitating a transition from a sparse skyrmion gas to a dense skyrmion fluid. Conversely, tensile strain drives a transition from skyrmions to antiskyrmions, as evidenced by the reversal of the system’s chirality. This demonstrates how strain mediates the competition
between DMI and other magnetic couplings. Specifically, here the competition between  chiral spin liquid physics and a $q=0$ order is crucial to the formation of these antiskyrmion-like textures at intermediate temperatures. 

Beyond the skyrmion lattice and fluid phases, strain exerts notable effects on other magnetic phases in the phase diagram.  For instance, in the FP phase, we find that at intermediate temperatures,  for strong enough negative strain, there is a sharp transition from the chiral spin liquid to an antiskyrmion phase, which is found for a small range of temperatures and upon cooling the system goes into the $q=0$ ordering. Similarly, in the helical phase, strain also induces an antiskyrmion phase characterized by elongated textures that populate the lattice at intermediate temperatures. 

The qualitative strain-dependent behavior we report is consistent with recent experimental findings of uniaxial strain driving skyrmion phase transitions in metallic multilayers \cite{ding2025multistep}, suggesting that similar mechanisms may be accessible in real materials.

Altogether, our study highlights strain as an important mechanism to manipulate magnetic properties in kagome-lattice systems. Strain not only stabilizes and reshapes skyrmion textures but also drives transitions between different magnetic states. Future work could build on these findings by addressing dynamical aspects, such as the motion and diffusion of skyrmions under strain,  which has been shown to change with the skyrmion density \cite{Raab2024,Zhang2023,Zhang2025}  , or by exploring experimental realizations designed to exploit strain-engineered chiral magnetism. External strain can be applied using a variety of methods, some at low strain \cite{kong2023direct}. Even though the strain amplitudes used in our simulations extend beyond what is usually achievable through epitaxial or substrate-induced deformation, comparable effects should already arise under experimentally accessible strain generated by controlled bending, or applied tensile stress, with methods similar to the ones used in Ref.\cite{ding2025multistep}.  Along this line, out-of-plane strain induced deformations could also be explored in future works, especially in connection with the control of the skyrmion,  and the stabilization of antiskyrmions. 

 We also note that the present study is restricted to a two-dimensional magnetic monolayer. The interplay between strain and long-range dipolar interactions in bulk or multilayer systems remains an interesting direction for future work.

By understanding how strain modifies skyrmion properties, we gain insights into using these topological spin textures in technological applications. The ability to tune skyrmion configurations, stabilize topological phases, and control skyrmion density offers exciting possibilities for the design of next-generation spintronic devices. In particular, our results indicate that strain engineering can enhance the thermal stability of skyrmions, enable skyrmion–antiskyrmion switching, and provide a practical, technologically relevant strategy for mechanically programmable spintronic devices.

\section*{Acknowledgments} 

F. A. G. A. and H. D. R. are partially supported by CONICET (PIP 2021-112200200101480CO),  SECyT UNLP PI+D  X947 and  Agencia I+D+i (PICT-2020-SERIEA-03205). F. A. G. A. acknowledges support from PIBAA 2872021010 0698CO (CONICET). G.D.S. and E.M.B thank support from SIIP-UNCuyo 06/M008-T1 grant, and CONICET PIP 2021-2023 11220200102578CO grant. This work used the TOKO Cluster from FCEN-UNCuyo and computational resources from UNC Superc cómputo (CCAD), which are part of the SNCAD-MinCyT, Argentina.


\begin{thebibliography}{50}%
\makeatletter
\providecommand \@ifxundefined [1]{%
 \@ifx{#1\undefined}
}%
\providecommand \@ifnum [1]{%
 \ifnum #1\expandafter \@firstoftwo
 \else \expandafter \@secondoftwo
 \fi
}%
\providecommand \@ifx [1]{%
 \ifx #1\expandafter \@firstoftwo
 \else \expandafter \@secondoftwo
 \fi
}%
\providecommand \natexlab [1]{#1}%
\providecommand \enquote  [1]{``#1''}%
\providecommand \bibnamefont  [1]{#1}%
\providecommand \bibfnamefont [1]{#1}%
\providecommand \citenamefont [1]{#1}%
\providecommand \href@noop [0]{\@secondoftwo}%
\providecommand \href [0]{\begingroup \@sanitize@url \@href}%
\providecommand \@href[1]{\@@startlink{#1}\@@href}%
\providecommand \@@href[1]{\endgroup#1\@@endlink}%
\providecommand \@sanitize@url [0]{\catcode `\\12\catcode `\$12\catcode
  `\&12\catcode `\#12\catcode `\^12\catcode `\_12\catcode `\%12\relax}%
\providecommand \@@startlink[1]{}%
\providecommand \@@endlink[0]{}%
\providecommand \url  [0]{\begingroup\@sanitize@url \@url }%
\providecommand \@url [1]{\endgroup\@href {#1}{\urlprefix }}%
\providecommand \urlprefix  [0]{URL }%
\providecommand \Eprint [0]{\href }%
\providecommand \doibase [0]{http://dx.doi.org/}%
\providecommand \selectlanguage [0]{\@gobble}%
\providecommand \bibinfo  [0]{\@secondoftwo}%
\providecommand \bibfield  [0]{\@secondoftwo}%
\providecommand \translation [1]{[#1]}%
\providecommand \BibitemOpen [0]{}%
\providecommand \bibitemStop [0]{}%
\providecommand \bibitemNoStop [0]{.\EOS\space}%
\providecommand \EOS [0]{\spacefactor3000\relax}%
\providecommand \BibitemShut  [1]{\csname bibitem#1\endcsname}%
\let\auto@bib@innerbib\@empty
\bibitem [{\citenamefont {Raab}\ \emph {et~al.}(2022)\citenamefont {Raab},
  \citenamefont {Brems}, \citenamefont {Beneke}, \citenamefont {Dohi},
  \citenamefont {Roth{\"o}rl}, \citenamefont {Kammerbauer}, \citenamefont
  {Mentink},\ and\ \citenamefont {Kl{\"a}ui}}]{raab2022brownian}%
  \BibitemOpen
  \bibfield  {author} {\bibinfo {author} {\bibfnamefont {Klaus}\ \bibnamefont
  {Raab}}, \bibinfo {author} {\bibfnamefont {Maarten~A}\ \bibnamefont {Brems}},
  \bibinfo {author} {\bibfnamefont {Grischa}\ \bibnamefont {Beneke}}, \bibinfo
  {author} {\bibfnamefont {Takaaki}\ \bibnamefont {Dohi}}, \bibinfo {author}
  {\bibfnamefont {Jan}\ \bibnamefont {Roth{\"o}rl}}, \bibinfo {author}
  {\bibfnamefont {Fabian}\ \bibnamefont {Kammerbauer}}, \bibinfo {author}
  {\bibfnamefont {Johan~H}\ \bibnamefont {Mentink}}, \ and\ \bibinfo {author}
  {\bibfnamefont {Mathias}\ \bibnamefont {Kl{\"a}ui}},\ }\bibfield  {title}
  {\enquote {\bibinfo {title} {Brownian reservoir computing realized using
  geometrically confined skyrmion dynamics},}\ }\href
  {https://doi.org/10.1038/s41467-022-34309-2} {\bibfield  {journal} {\bibinfo
  {journal} {Nature Communications}\ }\textbf {\bibinfo {volume} {13}},\
  \bibinfo {pages} {6982} (\bibinfo {year} {2022})}\BibitemShut {NoStop}%
\bibitem [{\citenamefont {Zhang}\ \emph {et~al.}(2015)\citenamefont {Zhang},
  \citenamefont {Ezawa},\ and\ \citenamefont {Zhou}}]{zhang2015magnetic}%
  \BibitemOpen
  \bibfield  {author} {\bibinfo {author} {\bibfnamefont {Xichao}\ \bibnamefont
  {Zhang}}, \bibinfo {author} {\bibfnamefont {Motohiko}\ \bibnamefont {Ezawa}},
  \ and\ \bibinfo {author} {\bibfnamefont {Yan}\ \bibnamefont {Zhou}},\
  }\bibfield  {title} {\enquote {\bibinfo {title} {Magnetic skyrmion logic
  gates: conversion, duplication and merging of skyrmions},}\ }\href
  {https://doi.org/10.1038/srep09400} {\bibfield  {journal} {\bibinfo
  {journal} {Scientific reports}\ }\textbf {\bibinfo {volume} {5}},\ \bibinfo
  {pages} {1--8} (\bibinfo {year} {2015})}\BibitemShut {NoStop}%
\bibitem [{\citenamefont {Pinna}\ \emph {et~al.}(2018)\citenamefont {Pinna},
  \citenamefont {Abreu~Araujo}, \citenamefont {Kim}, \citenamefont {Cros},
  \citenamefont {Querlioz}, \citenamefont {Bessiere}, \citenamefont {Droulez},\
  and\ \citenamefont {Grollier}}]{pinna2018skyrmion}%
  \BibitemOpen
  \bibfield  {author} {\bibinfo {author} {\bibfnamefont {Daniele}\ \bibnamefont
  {Pinna}}, \bibinfo {author} {\bibfnamefont {Flavio}\ \bibnamefont
  {Abreu~Araujo}}, \bibinfo {author} {\bibfnamefont {J-V}\ \bibnamefont {Kim}},
  \bibinfo {author} {\bibfnamefont {Vincent}\ \bibnamefont {Cros}}, \bibinfo
  {author} {\bibfnamefont {Damien}\ \bibnamefont {Querlioz}}, \bibinfo {author}
  {\bibfnamefont {Pierre}\ \bibnamefont {Bessiere}}, \bibinfo {author}
  {\bibfnamefont {Jacques}\ \bibnamefont {Droulez}}, \ and\ \bibinfo {author}
  {\bibfnamefont {Julie}\ \bibnamefont {Grollier}},\ }\bibfield  {title}
  {\enquote {\bibinfo {title} {Skyrmion gas manipulation for probabilistic
  computing},}\ }\href {https://doi.org/10.1103/PhysRevApplied.9.064018}
  {\bibfield  {journal} {\bibinfo  {journal} {Physical Review Applied}\
  }\textbf {\bibinfo {volume} {9}},\ \bibinfo {pages} {064018} (\bibinfo {year}
  {2018})}\BibitemShut {NoStop}%
\bibitem [{\citenamefont {Lee}\ \emph {et~al.}(2023)\citenamefont {Lee},
  \citenamefont {Msiska}, \citenamefont {Brems}, \citenamefont {Kl{\"a}ui},
  \citenamefont {Kurebayashi},\ and\ \citenamefont
  {Everschor-Sitte}}]{lee2023perspective}%
  \BibitemOpen
  \bibfield  {author} {\bibinfo {author} {\bibfnamefont {Oscar}\ \bibnamefont
  {Lee}}, \bibinfo {author} {\bibfnamefont {Robin}\ \bibnamefont {Msiska}},
  \bibinfo {author} {\bibfnamefont {Maarten~A}\ \bibnamefont {Brems}}, \bibinfo
  {author} {\bibfnamefont {Mathias}\ \bibnamefont {Kl{\"a}ui}}, \bibinfo
  {author} {\bibfnamefont {Hidekazu}\ \bibnamefont {Kurebayashi}}, \ and\
  \bibinfo {author} {\bibfnamefont {Karin}\ \bibnamefont {Everschor-Sitte}},\
  }\bibfield  {title} {\enquote {\bibinfo {title} {Perspective on
  unconventional computing using magnetic skyrmions},}\ }\href
  {https://doi.org/10.1063/5.0148469} {\bibfield  {journal} {\bibinfo
  {journal} {Applied Physics Letters}\ }\textbf {\bibinfo {volume} {122}}
  (\bibinfo {year} {2023})}\BibitemShut {NoStop}%
\bibitem [{\citenamefont {Moriya}(1960)}]{Moriya1960}%
  \BibitemOpen
  \bibfield  {author} {\bibinfo {author} {\bibfnamefont {T\^oru}\ \bibnamefont
  {Moriya}},\ }\bibfield  {title} {\enquote {\bibinfo {title} {Anisotropic
  superexchange interaction and weak ferromagnetism},}\ }\href {\doibase
  10.1103/PhysRev.120.91} {\bibfield  {journal} {\bibinfo  {journal} {Phys.
  Rev.}\ }\textbf {\bibinfo {volume} {120}},\ \bibinfo {pages} {91--98}
  (\bibinfo {year} {1960})}\BibitemShut {NoStop}%
\bibitem [{\citenamefont {Dzyaloshinskii}(1964)}]{dzyaloshinskii1964}%
  \BibitemOpen
  \bibfield  {author} {\bibinfo {author} {\bibfnamefont {IE}~\bibnamefont
  {Dzyaloshinskii}},\ }\bibfield  {title} {\enquote {\bibinfo {title} {Theory
  of helicoidal structures in antiferromagnets. i. nonmetals},}\ }\href@noop {}
  {\bibfield  {journal} {\bibinfo  {journal} {Sov. Phys. JETP}\ }\textbf
  {\bibinfo {volume} {19}},\ \bibinfo {pages} {960--971} (\bibinfo {year}
  {1964})}\BibitemShut {NoStop}%
\bibitem [{\citenamefont {Yi}\ \emph {et~al.}(2009)\citenamefont {Yi},
  \citenamefont {Onoda}, \citenamefont {Nagaosa},\ and\ \citenamefont
  {Han}}]{yi2009skyrmions}%
  \BibitemOpen
  \bibfield  {author} {\bibinfo {author} {\bibfnamefont {Su~Do}\ \bibnamefont
  {Yi}}, \bibinfo {author} {\bibfnamefont {Shigeki}\ \bibnamefont {Onoda}},
  \bibinfo {author} {\bibfnamefont {Naoto}\ \bibnamefont {Nagaosa}}, \ and\
  \bibinfo {author} {\bibfnamefont {Jung~Hoon}\ \bibnamefont {Han}},\
  }\bibfield  {title} {\enquote {\bibinfo {title} {Skyrmions and anomalous hall
  effect in a dzyaloshinskii-moriya spiral magnet},}\ }\href
  {https://doi.org/10.1103/PhysRevB.80.054416} {\bibfield  {journal} {\bibinfo
  {journal} {Physical Review B-Condensed Matter and Materials Physics}\
  }\textbf {\bibinfo {volume} {80}},\ \bibinfo {pages} {054416} (\bibinfo
  {year} {2009})}\BibitemShut {NoStop}%
\bibitem [{\citenamefont {Gao}\ \emph {et~al.}(2020)\citenamefont {Gao},
  \citenamefont {Rosales}, \citenamefont {Gomez~Albarracin}, \citenamefont
  {Tsurkan}, \citenamefont {Kaur}, \citenamefont {Fennell}, \citenamefont
  {Steffens}, \citenamefont {Boehm}, \citenamefont {{\v{C}}erm{\'a}k},
  \citenamefont {Schneidewind} \emph {et~al.}}]{gao2020fractional}%
  \BibitemOpen
  \bibfield  {author} {\bibinfo {author} {\bibfnamefont {Shang}\ \bibnamefont
  {Gao}}, \bibinfo {author} {\bibfnamefont {H~Diego}\ \bibnamefont {Rosales}},
  \bibinfo {author} {\bibfnamefont {Flavia~A}\ \bibnamefont
  {Gomez~Albarracin}}, \bibinfo {author} {\bibfnamefont {Vladimir}\
  \bibnamefont {Tsurkan}}, \bibinfo {author} {\bibfnamefont {Guratinder}\
  \bibnamefont {Kaur}}, \bibinfo {author} {\bibfnamefont {Tom}\ \bibnamefont
  {Fennell}}, \bibinfo {author} {\bibfnamefont {Paul}\ \bibnamefont
  {Steffens}}, \bibinfo {author} {\bibfnamefont {Martin}\ \bibnamefont
  {Boehm}}, \bibinfo {author} {\bibfnamefont {Petr}\ \bibnamefont
  {{\v{C}}erm{\'a}k}}, \bibinfo {author} {\bibfnamefont {Astrid}\ \bibnamefont
  {Schneidewind}},  \emph {et~al.},\ }\bibfield  {title} {\enquote {\bibinfo
  {title} {Fractional antiferromagnetic skyrmion lattice induced by anisotropic
  couplings},}\ }\href {https://doi.org/10.1038/s41586-020-2716-8} {\bibfield
  {journal} {\bibinfo  {journal} {Nature}\ }\textbf {\bibinfo {volume} {586}},\
  \bibinfo {pages} {37--41} (\bibinfo {year} {2020})}\BibitemShut {NoStop}%
\bibitem [{\citenamefont {Amoroso}\ \emph {et~al.}(2020)\citenamefont
  {Amoroso}, \citenamefont {Barone},\ and\ \citenamefont
  {Picozzi}}]{amoroso2020spontaneous}%
  \BibitemOpen
  \bibfield  {author} {\bibinfo {author} {\bibfnamefont {Danila}\ \bibnamefont
  {Amoroso}}, \bibinfo {author} {\bibfnamefont {Paolo}\ \bibnamefont {Barone}},
  \ and\ \bibinfo {author} {\bibfnamefont {Silvia}\ \bibnamefont {Picozzi}},\
  }\bibfield  {title} {\enquote {\bibinfo {title} {Spontaneous skyrmionic
  lattice from anisotropic symmetric exchange in a ni-halide monolayer},}\
  }\href {https://doi.org/10.1038/s41467-020-19535-w} {\bibfield  {journal}
  {\bibinfo  {journal} {Nature Communications}\ }\textbf {\bibinfo {volume}
  {11}},\ \bibinfo {pages} {5784} (\bibinfo {year} {2020})}\BibitemShut
  {NoStop}%
\bibitem [{\citenamefont {Wang}\ \emph {et~al.}(2021)\citenamefont {Wang},
  \citenamefont {Su}, \citenamefont {Lin},\ and\ \citenamefont
  {Batista}}]{wang2021meron}%
  \BibitemOpen
  \bibfield  {author} {\bibinfo {author} {\bibfnamefont {Zhentao}\ \bibnamefont
  {Wang}}, \bibinfo {author} {\bibfnamefont {Ying}\ \bibnamefont {Su}},
  \bibinfo {author} {\bibfnamefont {Shi-Zeng}\ \bibnamefont {Lin}}, \ and\
  \bibinfo {author} {\bibfnamefont {Cristian~D}\ \bibnamefont {Batista}},\
  }\bibfield  {title} {\enquote {\bibinfo {title} {Meron, skyrmion, and vortex
  crystals in centrosymmetric tetragonal magnets},}\ }\href
  {https://doi.org/10.1103/PhysRevB.103.104408} {\bibfield  {journal} {\bibinfo
   {journal} {Physical Review B}\ }\textbf {\bibinfo {volume} {103}},\ \bibinfo
  {pages} {104408} (\bibinfo {year} {2021})}\BibitemShut {NoStop}%
\bibitem [{\citenamefont {Rosales}\ \emph {et~al.}(2022)\citenamefont
  {Rosales}, \citenamefont {Albarrac{\'\i}n}, \citenamefont {Guratinder},
  \citenamefont {Tsurkan}, \citenamefont {Prodan}, \citenamefont {Ressouche},\
  and\ \citenamefont {Zaharko}}]{rosales2022anisotropy}%
  \BibitemOpen
  \bibfield  {author} {\bibinfo {author} {\bibfnamefont {H~D}\ \bibnamefont
  {Rosales}}, \bibinfo {author} {\bibfnamefont {F~A~G{\'o}mez}\ \bibnamefont
  {Albarrac{\'\i}n}}, \bibinfo {author} {\bibfnamefont {K}~\bibnamefont
  {Guratinder}}, \bibinfo {author} {\bibfnamefont {Vladimir}\ \bibnamefont
  {Tsurkan}}, \bibinfo {author} {\bibfnamefont {Lilian}\ \bibnamefont
  {Prodan}}, \bibinfo {author} {\bibfnamefont {Eric}\ \bibnamefont
  {Ressouche}}, \ and\ \bibinfo {author} {\bibfnamefont {Oksana}\ \bibnamefont
  {Zaharko}},\ }\bibfield  {title} {\enquote {\bibinfo {title}
  {Anisotropy-driven response of the fractional antiferromagnetic skyrmion
  lattice in mnsc2s4 to applied magnetic fields},}\ }\href
  {https://doi.org/10.1103/PhysRevB.105.224402} {\bibfield  {journal} {\bibinfo
   {journal} {Physical Review B}\ }\textbf {\bibinfo {volume} {105}},\ \bibinfo
  {pages} {224402} (\bibinfo {year} {2022})}\BibitemShut {NoStop}%
\bibitem [{\citenamefont {Wang}\ \emph {et~al.}(2020)\citenamefont {Wang},
  \citenamefont {Su}, \citenamefont {Lin},\ and\ \citenamefont
  {Batista}}]{wang2020skyrmion}%
  \BibitemOpen
  \bibfield  {author} {\bibinfo {author} {\bibfnamefont {Zhentao}\ \bibnamefont
  {Wang}}, \bibinfo {author} {\bibfnamefont {Ying}\ \bibnamefont {Su}},
  \bibinfo {author} {\bibfnamefont {Shi-Zeng}\ \bibnamefont {Lin}}, \ and\
  \bibinfo {author} {\bibfnamefont {Cristian~D}\ \bibnamefont {Batista}},\
  }\bibfield  {title} {\enquote {\bibinfo {title} {Skyrmion crystal from rkky
  interaction mediated by 2d electron gas},}\ }\href
  {https://doi.org/10.1103/PhysRevLett.124.207201} {\bibfield  {journal}
  {\bibinfo  {journal} {Physical Review Letters}\ }\textbf {\bibinfo {volume}
  {124}},\ \bibinfo {pages} {207201} (\bibinfo {year} {2020})}\BibitemShut
  {NoStop}%
\bibitem [{\citenamefont {Utesov}(2021)}]{UtesovDipolar}%
  \BibitemOpen
  \bibfield  {author} {\bibinfo {author} {\bibfnamefont {Oleg~I.}\ \bibnamefont
  {Utesov}},\ }\bibfield  {title} {\enquote {\bibinfo {title}
  {Thermodynamically stable skyrmion lattice in a tetragonal frustrated
  antiferromagnet with dipolar interaction},}\ }\href {\doibase
  10.1103/PhysRevB.103.064414} {\bibfield  {journal} {\bibinfo  {journal}
  {Phys. Rev. B}\ }\textbf {\bibinfo {volume} {103}},\ \bibinfo {pages}
  {064414} (\bibinfo {year} {2021})}\BibitemShut {NoStop}%
\bibitem [{\citenamefont {Paul}\ \emph {et~al.}(2020)\citenamefont {Paul},
  \citenamefont {Haldar}, \citenamefont {Von~Malottki},\ and\ \citenamefont
  {Heinze}}]{paul2020role}%
  \BibitemOpen
  \bibfield  {author} {\bibinfo {author} {\bibfnamefont {Souvik}\ \bibnamefont
  {Paul}}, \bibinfo {author} {\bibfnamefont {Soumyajyoti}\ \bibnamefont
  {Haldar}}, \bibinfo {author} {\bibfnamefont {Stephan}\ \bibnamefont
  {Von~Malottki}}, \ and\ \bibinfo {author} {\bibfnamefont {Stefan}\
  \bibnamefont {Heinze}},\ }\bibfield  {title} {\enquote {\bibinfo {title}
  {Role of higher-order exchange interactions for skyrmion stability},}\ }\href
  {https://doi.org/10.1038/s41467-020-18473-x} {\bibfield  {journal} {\bibinfo
  {journal} {Nature Communications}\ }\textbf {\bibinfo {volume} {11}},\
  \bibinfo {pages} {4756} (\bibinfo {year} {2020})}\BibitemShut {NoStop}%
\bibitem [{\citenamefont {Okubo}\ \emph {et~al.}(2012)\citenamefont {Okubo},
  \citenamefont {Chung},\ and\ \citenamefont {Kawamura}}]{okubo2012multiple}%
  \BibitemOpen
  \bibfield  {author} {\bibinfo {author} {\bibfnamefont {Tsuyoshi}\
  \bibnamefont {Okubo}}, \bibinfo {author} {\bibfnamefont {Sungki}\
  \bibnamefont {Chung}}, \ and\ \bibinfo {author} {\bibfnamefont {Hikaru}\
  \bibnamefont {Kawamura}},\ }\bibfield  {title} {\enquote {\bibinfo {title}
  {Multiple-q states and the skyrmion lattice of the triangular-lattice
  heisenberg antiferromagnet under magnetic fields},}\ }\href
  {https://doi.org/10.1103/PhysRevLett.108.017206} {\bibfield  {journal}
  {\bibinfo  {journal} {Physical Review Letters}\ }\textbf {\bibinfo {volume}
  {108}},\ \bibinfo {pages} {017206} (\bibinfo {year} {2012})}\BibitemShut
  {NoStop}%
\bibitem [{\citenamefont {Mohylna}\ \emph {et~al.}(2022)\citenamefont
  {Mohylna}, \citenamefont {Albarrac{\'\i}n}, \citenamefont
  {{\v{Z}}ukovi{\v{c}}},\ and\ \citenamefont
  {Rosales}}]{mohylna2022spontaneous}%
  \BibitemOpen
  \bibfield  {author} {\bibinfo {author} {\bibfnamefont {M}~\bibnamefont
  {Mohylna}}, \bibinfo {author} {\bibfnamefont {F~A~G{\'o}mez}\ \bibnamefont
  {Albarrac{\'\i}n}}, \bibinfo {author} {\bibfnamefont {M}~\bibnamefont
  {{\v{Z}}ukovi{\v{c}}}}, \ and\ \bibinfo {author} {\bibfnamefont {H~D}\
  \bibnamefont {Rosales}},\ }\bibfield  {title} {\enquote {\bibinfo {title}
  {Spontaneous antiferromagnetic skyrmion/antiskyrmion lattice and spiral
  spin-liquid states in the frustrated triangular lattice},}\ }\href
  {https://doi.org/10.1103/PhysRevB.106.224406} {\bibfield  {journal} {\bibinfo
   {journal} {Physical Review B}\ }\textbf {\bibinfo {volume} {106}},\ \bibinfo
  {pages} {224406} (\bibinfo {year} {2022})}\BibitemShut {NoStop}%
\bibitem [{\citenamefont {Rosales}\ \emph {et~al.}(2023)\citenamefont
  {Rosales}, \citenamefont {Albarrac{\'\i}n}, \citenamefont {Pujol},\ and\
  \citenamefont {Jaubert}}]{rosales2023skyrmion}%
  \BibitemOpen
  \bibfield  {author} {\bibinfo {author} {\bibfnamefont {H~Diego}\ \bibnamefont
  {Rosales}}, \bibinfo {author} {\bibfnamefont {Flavia A~G{\'o}mez}\
  \bibnamefont {Albarrac{\'\i}n}}, \bibinfo {author} {\bibfnamefont {Pierre}\
  \bibnamefont {Pujol}}, \ and\ \bibinfo {author} {\bibfnamefont {Ludovic~DC}\
  \bibnamefont {Jaubert}},\ }\bibfield  {title} {\enquote {\bibinfo {title}
  {Skyrmion fluid and bimeron glass protected by a chiral spin liquid on a
  kagome lattice},}\ }\href {https://doi.org/10.1103/PhysRevLett.130.106703}
  {\bibfield  {journal} {\bibinfo  {journal} {Physical Review Letters}\
  }\textbf {\bibinfo {volume} {130}},\ \bibinfo {pages} {106703} (\bibinfo
  {year} {2023})}\BibitemShut {NoStop}%
\bibitem [{\citenamefont {G{\'o}mez~Albarrac{\'\i}n}\ \emph
  {et~al.}(2024)\citenamefont {G{\'o}mez~Albarrac{\'\i}n}, \citenamefont
  {Rosales}, \citenamefont {Udagawa}, \citenamefont {Pujol},\ and\
  \citenamefont {Jaubert}}]{gomez2024chiral}%
  \BibitemOpen
  \bibfield  {author} {\bibinfo {author} {\bibfnamefont {F~A}\ \bibnamefont
  {G{\'o}mez~Albarrac{\'\i}n}}, \bibinfo {author} {\bibfnamefont {H~Diego}\
  \bibnamefont {Rosales}}, \bibinfo {author} {\bibfnamefont {Masafumi}\
  \bibnamefont {Udagawa}}, \bibinfo {author} {\bibfnamefont {P}~\bibnamefont
  {Pujol}}, \ and\ \bibinfo {author} {\bibfnamefont {Ludovic~DC}\ \bibnamefont
  {Jaubert}},\ }\bibfield  {title} {\enquote {\bibinfo {title} {From chiral
  spin liquids to skyrmion fluids and crystals, and their interplay with
  itinerant electrons},}\ }\href {https://doi.org/10.1103/PhysRevB.109.064426}
  {\bibfield  {journal} {\bibinfo  {journal} {Physical Review B}\ }\textbf
  {\bibinfo {volume} {109}},\ \bibinfo {pages} {064426} (\bibinfo {year}
  {2024})}\BibitemShut {NoStop}%
\bibitem [{\citenamefont {Nii}\ \emph {et~al.}(2015)\citenamefont {Nii},
  \citenamefont {Nakajima}, \citenamefont {Kikkawa}, \citenamefont {Yamasaki},
  \citenamefont {Ohishi}, \citenamefont {Suzuki}, \citenamefont {Taguchi},
  \citenamefont {Arima}, \citenamefont {Tokura},\ and\ \citenamefont
  {Iwasa}}]{nii2015uniaxial}%
  \BibitemOpen
  \bibfield  {author} {\bibinfo {author} {\bibfnamefont {Y}~\bibnamefont
  {Nii}}, \bibinfo {author} {\bibfnamefont {T}~\bibnamefont {Nakajima}},
  \bibinfo {author} {\bibfnamefont {A}~\bibnamefont {Kikkawa}}, \bibinfo
  {author} {\bibfnamefont {Y}~\bibnamefont {Yamasaki}}, \bibinfo {author}
  {\bibfnamefont {K}~\bibnamefont {Ohishi}}, \bibinfo {author} {\bibfnamefont
  {J}~\bibnamefont {Suzuki}}, \bibinfo {author} {\bibfnamefont {Y}~\bibnamefont
  {Taguchi}}, \bibinfo {author} {\bibfnamefont {T}~\bibnamefont {Arima}},
  \bibinfo {author} {\bibfnamefont {Y}~\bibnamefont {Tokura}}, \ and\ \bibinfo
  {author} {\bibfnamefont {Y}~\bibnamefont {Iwasa}},\ }\bibfield  {title}
  {\enquote {\bibinfo {title} {Uniaxial stress control of skyrmion phase},}\
  }\href {https://doi.org/10.1038/ncomms9539} {\bibfield  {journal} {\bibinfo
  {journal} {Nature communications}\ }\textbf {\bibinfo {volume} {6}},\
  \bibinfo {pages} {8539} (\bibinfo {year} {2015})}\BibitemShut {NoStop}%
\bibitem [{\citenamefont {Shibata}\ \emph {et~al.}(2015)\citenamefont
  {Shibata}, \citenamefont {Iwasaki}, \citenamefont {Kanazawa}, \citenamefont
  {Aizawa}, \citenamefont {Tanigaki}, \citenamefont {Shirai}, \citenamefont
  {Nakajima}, \citenamefont {Kubota}, \citenamefont {Kawasaki}, \citenamefont
  {Park} \emph {et~al.}}]{shibata2015large}%
  \BibitemOpen
  \bibfield  {author} {\bibinfo {author} {\bibfnamefont {K}~\bibnamefont
  {Shibata}}, \bibinfo {author} {\bibfnamefont {J}~\bibnamefont {Iwasaki}},
  \bibinfo {author} {\bibfnamefont {N}~\bibnamefont {Kanazawa}}, \bibinfo
  {author} {\bibfnamefont {S}~\bibnamefont {Aizawa}}, \bibinfo {author}
  {\bibfnamefont {T}~\bibnamefont {Tanigaki}}, \bibinfo {author} {\bibfnamefont
  {M}~\bibnamefont {Shirai}}, \bibinfo {author} {\bibfnamefont {T}~\bibnamefont
  {Nakajima}}, \bibinfo {author} {\bibfnamefont {M}~\bibnamefont {Kubota}},
  \bibinfo {author} {\bibfnamefont {M}~\bibnamefont {Kawasaki}}, \bibinfo
  {author} {\bibfnamefont {HS}~\bibnamefont {Park}},  \emph {et~al.},\
  }\bibfield  {title} {\enquote {\bibinfo {title} {Large anisotropic
  deformation of skyrmions in strained crystal},}\ }\href
  {https://doi.org/10.1038/nnano.2015.113} {\bibfield  {journal} {\bibinfo
  {journal} {Nature nanotechnology}\ }\textbf {\bibinfo {volume} {10}},\
  \bibinfo {pages} {589--592} (\bibinfo {year} {2015})}\BibitemShut {NoStop}%
\bibitem [{\citenamefont {Camosi}\ \emph {et~al.}(2017)\citenamefont {Camosi},
  \citenamefont {Rohart}, \citenamefont {Fruchart}, \citenamefont {Pizzini},
  \citenamefont {Belmeguenai}, \citenamefont {Roussign{\'e}}, \citenamefont
  {Stashkevich}, \citenamefont {Cherif}, \citenamefont {Ranno}, \citenamefont
  {De~Santis} \emph {et~al.}}]{camosi2017anisotropic}%
  \BibitemOpen
  \bibfield  {author} {\bibinfo {author} {\bibfnamefont {Lorenzo}\ \bibnamefont
  {Camosi}}, \bibinfo {author} {\bibfnamefont {Stanislas}\ \bibnamefont
  {Rohart}}, \bibinfo {author} {\bibfnamefont {Olivier}\ \bibnamefont
  {Fruchart}}, \bibinfo {author} {\bibfnamefont {Stefania}\ \bibnamefont
  {Pizzini}}, \bibinfo {author} {\bibfnamefont {Mohamed}\ \bibnamefont
  {Belmeguenai}}, \bibinfo {author} {\bibfnamefont {Yves}\ \bibnamefont
  {Roussign{\'e}}}, \bibinfo {author} {\bibfnamefont {Andre{\"\i}}\
  \bibnamefont {Stashkevich}}, \bibinfo {author} {\bibfnamefont {Salim~Mourad}\
  \bibnamefont {Cherif}}, \bibinfo {author} {\bibfnamefont {Laurent}\
  \bibnamefont {Ranno}}, \bibinfo {author} {\bibfnamefont {Maurizio}\
  \bibnamefont {De~Santis}},  \emph {et~al.},\ }\bibfield  {title} {\enquote
  {\bibinfo {title} {Anisotropic dzyaloshinskii-moriya interaction in ultrathin
  epitaxial au/co/w (110)},}\ }\href
  {https://doi.org/10.1103/PhysRevB.95.214422} {\bibfield  {journal} {\bibinfo
  {journal} {Physical Review B}\ }\textbf {\bibinfo {volume} {95}},\ \bibinfo
  {pages} {214422} (\bibinfo {year} {2017})}\BibitemShut {NoStop}%
\bibitem [{\citenamefont {Deger}(2020)}]{deger2020strain}%
  \BibitemOpen
  \bibfield  {author} {\bibinfo {author} {\bibfnamefont {Caner}\ \bibnamefont
  {Deger}},\ }\bibfield  {title} {\enquote {\bibinfo {title} {Strain-enhanced
  dzyaloshinskii--moriya interaction at co/pt interfaces},}\ }\href
  {https://doi.org/10.1038/s41598-020-69360-w} {\bibfield  {journal} {\bibinfo
  {journal} {Scientific reports}\ }\textbf {\bibinfo {volume} {10}},\ \bibinfo
  {pages} {12314} (\bibinfo {year} {2020})}\BibitemShut {NoStop}%
\bibitem [{\citenamefont {Gusev}\ \emph {et~al.}(2020)\citenamefont {Gusev},
  \citenamefont {Sadovnikov}, \citenamefont {Nikitov}, \citenamefont
  {Sapozhnikov},\ and\ \citenamefont {Udalov}}]{gusev2020manipulation}%
  \BibitemOpen
  \bibfield  {author} {\bibinfo {author} {\bibfnamefont {NS}~\bibnamefont
  {Gusev}}, \bibinfo {author} {\bibfnamefont {AV}~\bibnamefont {Sadovnikov}},
  \bibinfo {author} {\bibfnamefont {SA}~\bibnamefont {Nikitov}}, \bibinfo
  {author} {\bibfnamefont {MV}~\bibnamefont {Sapozhnikov}}, \ and\ \bibinfo
  {author} {\bibfnamefont {OG}~\bibnamefont {Udalov}},\ }\bibfield  {title}
  {\enquote {\bibinfo {title} {Manipulation of the dzyaloshinskii--moriya
  interaction in co/pt multilayers with strain},}\ }\href
  {https://doi.org/10.1103/PhysRevLett.124.157202} {\bibfield  {journal}
  {\bibinfo  {journal} {Physical review letters}\ }\textbf {\bibinfo {volume}
  {124}},\ \bibinfo {pages} {157202} (\bibinfo {year} {2020})}\BibitemShut
  {NoStop}%
\bibitem [{\citenamefont {Tanaka}\ \emph {et~al.}(2020)\citenamefont {Tanaka},
  \citenamefont {Sugawara},\ and\ \citenamefont
  {Mochizuki}}]{tanaka2020theoretical}%
  \BibitemOpen
  \bibfield  {author} {\bibinfo {author} {\bibfnamefont {Kohei}\ \bibnamefont
  {Tanaka}}, \bibinfo {author} {\bibfnamefont {Ryosuke}\ \bibnamefont
  {Sugawara}}, \ and\ \bibinfo {author} {\bibfnamefont {Masahito}\ \bibnamefont
  {Mochizuki}},\ }\bibfield  {title} {\enquote {\bibinfo {title} {Theoretical
  study on stabilization and destabilization of magnetic skyrmions by
  uniaxial-strain-induced anisotropic dzyaloshinskii-moriya interactions},}\
  }\href {https://doi.org/10.1103/PhysRevMaterials.4.034404} {\bibfield
  {journal} {\bibinfo  {journal} {Physical Review Materials}\ }\textbf
  {\bibinfo {volume} {4}},\ \bibinfo {pages} {034404} (\bibinfo {year}
  {2020})}\BibitemShut {NoStop}%
\bibitem [{\citenamefont {Feng}\ \emph {et~al.}(2021)\citenamefont {Feng},
  \citenamefont {Meng}, \citenamefont {Wang}, \citenamefont {Jiang},
  \citenamefont {Mehmood}, \citenamefont {Cao}, \citenamefont {Lv},
  \citenamefont {Yang}, \citenamefont {Wang}, \citenamefont {Zhao} \emph
  {et~al.}}]{feng2021field}%
  \BibitemOpen
  \bibfield  {author} {\bibinfo {author} {\bibfnamefont {Chun}\ \bibnamefont
  {Feng}}, \bibinfo {author} {\bibfnamefont {Fei}\ \bibnamefont {Meng}},
  \bibinfo {author} {\bibfnamefont {Yadong}\ \bibnamefont {Wang}}, \bibinfo
  {author} {\bibfnamefont {Jiawei}\ \bibnamefont {Jiang}}, \bibinfo {author}
  {\bibfnamefont {Nasir}\ \bibnamefont {Mehmood}}, \bibinfo {author}
  {\bibfnamefont {Yi}~\bibnamefont {Cao}}, \bibinfo {author} {\bibfnamefont
  {Xiaowei}\ \bibnamefont {Lv}}, \bibinfo {author} {\bibfnamefont {Feng}\
  \bibnamefont {Yang}}, \bibinfo {author} {\bibfnamefont {Lei}\ \bibnamefont
  {Wang}}, \bibinfo {author} {\bibfnamefont {Yongkang}\ \bibnamefont {Zhao}},
  \emph {et~al.},\ }\bibfield  {title} {\enquote {\bibinfo {title} {Field-free
  manipulation of skyrmion creation and annihilation by tunable strain
  engineering},}\ }\href {https://doi.org/10.1002/adfm.202008715} {\bibfield
  {journal} {\bibinfo  {journal} {Advanced Functional Materials}\ }\textbf
  {\bibinfo {volume} {31}},\ \bibinfo {pages} {2008715} (\bibinfo {year}
  {2021})}\BibitemShut {NoStop}%
\bibitem [{\citenamefont {Zhang}\ \emph {et~al.}(2021)\citenamefont {Zhang},
  \citenamefont {Liu}, \citenamefont {Dong}, \citenamefont {Wu}, \citenamefont
  {Zhang}, \citenamefont {Wang}, \citenamefont {Lu}, \citenamefont
  {R{\"u}ckriegel}, \citenamefont {Wang}, \citenamefont {Duine} \emph
  {et~al.}}]{zhang2021strain}%
  \BibitemOpen
  \bibfield  {author} {\bibinfo {author} {\bibfnamefont {Yuelin}\ \bibnamefont
  {Zhang}}, \bibinfo {author} {\bibfnamefont {Jie}\ \bibnamefont {Liu}},
  \bibinfo {author} {\bibfnamefont {Yongqi}\ \bibnamefont {Dong}}, \bibinfo
  {author} {\bibfnamefont {Shizhe}\ \bibnamefont {Wu}}, \bibinfo {author}
  {\bibfnamefont {Jianyu}\ \bibnamefont {Zhang}}, \bibinfo {author}
  {\bibfnamefont {Jie}\ \bibnamefont {Wang}}, \bibinfo {author} {\bibfnamefont
  {Jingdi}\ \bibnamefont {Lu}}, \bibinfo {author} {\bibfnamefont {Andreas}\
  \bibnamefont {R{\"u}ckriegel}}, \bibinfo {author} {\bibfnamefont {Hanchen}\
  \bibnamefont {Wang}}, \bibinfo {author} {\bibfnamefont {Rembert}\
  \bibnamefont {Duine}},  \emph {et~al.},\ }\bibfield  {title} {\enquote
  {\bibinfo {title} {Strain-driven dzyaloshinskii-moriya interaction for
  room-temperature magnetic skyrmions},}\ }\href
  {https://doi.org/10.1103/PhysRevLett.127.117204} {\bibfield  {journal}
  {\bibinfo  {journal} {Physical Review Letters}\ }\textbf {\bibinfo {volume}
  {127}},\ \bibinfo {pages} {117204} (\bibinfo {year} {2021})}\BibitemShut
  {NoStop}%
\bibitem [{\citenamefont {El~Hog}\ \emph {et~al.}(2022)\citenamefont {El~Hog},
  \citenamefont {Kato}, \citenamefont {Hongo}, \citenamefont {Koibuchi},
  \citenamefont {Diguet}, \citenamefont {Uchimoto},\ and\ \citenamefont
  {Diep}}]{el2022stability}%
  \BibitemOpen
  \bibfield  {author} {\bibinfo {author} {\bibfnamefont {Sahbi}\ \bibnamefont
  {El~Hog}}, \bibinfo {author} {\bibfnamefont {Fumitake}\ \bibnamefont {Kato}},
  \bibinfo {author} {\bibfnamefont {Satoshi}\ \bibnamefont {Hongo}}, \bibinfo
  {author} {\bibfnamefont {Hiroshi}\ \bibnamefont {Koibuchi}}, \bibinfo
  {author} {\bibfnamefont {Gildas}\ \bibnamefont {Diguet}}, \bibinfo {author}
  {\bibfnamefont {Tetsuya}\ \bibnamefont {Uchimoto}}, \ and\ \bibinfo {author}
  {\bibfnamefont {Hung~T}\ \bibnamefont {Diep}},\ }\bibfield  {title} {\enquote
  {\bibinfo {title} {The stability of 3d skyrmions under mechanical stress
  studied via monte carlo calculations},}\ }\href
  {https://doi.org/10.1016/j.rinp.2022.105578} {\bibfield  {journal} {\bibinfo
  {journal} {Results in Physics}\ }\textbf {\bibinfo {volume} {38}},\ \bibinfo
  {pages} {105578} (\bibinfo {year} {2022})}\BibitemShut {NoStop}%
\bibitem [{\citenamefont {Li}\ \emph {et~al.}(2022)\citenamefont {Li},
  \citenamefont {Haldar},\ and\ \citenamefont {Heinze}}]{li2022strain}%
  \BibitemOpen
  \bibfield  {author} {\bibinfo {author} {\bibfnamefont {Dongzhe}\ \bibnamefont
  {Li}}, \bibinfo {author} {\bibfnamefont {Soumyajyoti}\ \bibnamefont
  {Haldar}}, \ and\ \bibinfo {author} {\bibfnamefont {Stefan}\ \bibnamefont
  {Heinze}},\ }\bibfield  {title} {\enquote {\bibinfo {title} {Strain-driven
  zero-field near-10 nm skyrmions in two-dimensional van der waals
  heterostructures},}\ }\href {https://doi.org/10.1021/acs.nanolett.2c03287}
  {\bibfield  {journal} {\bibinfo  {journal} {Nano Letters}\ }\textbf {\bibinfo
  {volume} {22}},\ \bibinfo {pages} {7706--7713} (\bibinfo {year}
  {2022})}\BibitemShut {NoStop}%
\bibitem [{\citenamefont {Littlehales}\ \emph {et~al.}(2022)\citenamefont
  {Littlehales}, \citenamefont {Turnbull}, \citenamefont {Wilson},
  \citenamefont {Birch}, \citenamefont {Popescu}, \citenamefont {Jaouen},
  \citenamefont {Verezhak}, \citenamefont {Balakrishnan},\ and\ \citenamefont
  {Hatton}}]{littlehales2022enhanced}%
  \BibitemOpen
  \bibfield  {author} {\bibinfo {author} {\bibfnamefont {MT}~\bibnamefont
  {Littlehales}}, \bibinfo {author} {\bibfnamefont {LA}~\bibnamefont
  {Turnbull}}, \bibinfo {author} {\bibfnamefont {MN}~\bibnamefont {Wilson}},
  \bibinfo {author} {\bibfnamefont {MT}~\bibnamefont {Birch}}, \bibinfo
  {author} {\bibfnamefont {H}~\bibnamefont {Popescu}}, \bibinfo {author}
  {\bibfnamefont {N}~\bibnamefont {Jaouen}}, \bibinfo {author} {\bibfnamefont
  {JAT}\ \bibnamefont {Verezhak}}, \bibinfo {author} {\bibfnamefont
  {G}~\bibnamefont {Balakrishnan}}, \ and\ \bibinfo {author} {\bibfnamefont
  {PD}~\bibnamefont {Hatton}},\ }\bibfield  {title} {\enquote {\bibinfo {title}
  {Enhanced skyrmion metastability under applied strain in fege},}\ }\href
  {https://doi.org/10.1103/PhysRevB.106.214434} {\bibfield  {journal} {\bibinfo
   {journal} {Physical Review B}\ }\textbf {\bibinfo {volume} {106}},\ \bibinfo
  {pages} {214434} (\bibinfo {year} {2022})}\BibitemShut {NoStop}%
\bibitem [{\citenamefont {Dong}\ \emph {et~al.}(2023)\citenamefont {Dong},
  \citenamefont {Wang}, \citenamefont {Shi}, \citenamefont {Liang},
  \citenamefont {Jafri}, \citenamefont {Hu}, \citenamefont {Jin},\ and\
  \citenamefont {Huang}}]{dong2023strain}%
  \BibitemOpen
  \bibfield  {author} {\bibinfo {author} {\bibfnamefont {Shouzhe}\ \bibnamefont
  {Dong}}, \bibinfo {author} {\bibfnamefont {Jing}\ \bibnamefont {Wang}},
  \bibinfo {author} {\bibfnamefont {Xiaoming}\ \bibnamefont {Shi}}, \bibinfo
  {author} {\bibfnamefont {Deshan}\ \bibnamefont {Liang}}, \bibinfo {author}
  {\bibfnamefont {Hasnain~Mehdi}\ \bibnamefont {Jafri}}, \bibinfo {author}
  {\bibfnamefont {Chengchao}\ \bibnamefont {Hu}}, \bibinfo {author}
  {\bibfnamefont {Ke}~\bibnamefont {Jin}}, \ and\ \bibinfo {author}
  {\bibfnamefont {Houbing}\ \bibnamefont {Huang}},\ }\bibfield  {title}
  {\enquote {\bibinfo {title} {Strain-tuning bloch-and n{\'e}el-type magnetic
  skyrmions: A phase-field simulation},}\ }\href
  {https://doi.org/10.1016/j.scriptamat.2022.114994} {\bibfield  {journal}
  {\bibinfo  {journal} {Scripta Materialia}\ }\textbf {\bibinfo {volume}
  {222}},\ \bibinfo {pages} {114994} (\bibinfo {year} {2023})}\BibitemShut
  {NoStop}%
\bibitem [{\citenamefont {Mito}\ \emph {et~al.}(2024)\citenamefont {Mito},
  \citenamefont {Tajiri}, \citenamefont {Kousaka}, \citenamefont {Miyagawa},
  \citenamefont {Koyama}, \citenamefont {Akimitsu},\ and\ \citenamefont
  {Inoue}}]{mito2024magnetostriction}%
  \BibitemOpen
  \bibfield  {author} {\bibinfo {author} {\bibfnamefont {Masaki}\ \bibnamefont
  {Mito}}, \bibinfo {author} {\bibfnamefont {Takayuki}\ \bibnamefont {Tajiri}},
  \bibinfo {author} {\bibfnamefont {Yusuke}\ \bibnamefont {Kousaka}}, \bibinfo
  {author} {\bibfnamefont {Marina}\ \bibnamefont {Miyagawa}}, \bibinfo {author}
  {\bibfnamefont {Tamami}\ \bibnamefont {Koyama}}, \bibinfo {author}
  {\bibfnamefont {Jun}\ \bibnamefont {Akimitsu}}, \ and\ \bibinfo {author}
  {\bibfnamefont {Katsuya}\ \bibnamefont {Inoue}},\ }\bibfield  {title}
  {\enquote {\bibinfo {title} {Magnetostriction related to skyrmion-lattice
  formation in chiral magnet fege},}\ }\href
  {https://doi.org/10.1063/5.0227382} {\bibfield  {journal} {\bibinfo
  {journal} {Journal of Applied Physics}\ }\textbf {\bibinfo {volume} {136}}
  (\bibinfo {year} {2024})}\BibitemShut {NoStop}%
\bibitem [{\citenamefont {Yang}\ \emph {et~al.}(2024)\citenamefont {Yang},
  \citenamefont {Dou}, \citenamefont {Li}, \citenamefont {Dai}, \citenamefont
  {Huang},\ and\ \citenamefont {Ma}}]{yang2024strain}%
  \BibitemOpen
  \bibfield  {author} {\bibinfo {author} {\bibfnamefont {Junhuang}\
  \bibnamefont {Yang}}, \bibinfo {author} {\bibfnamefont {Kaiying}\
  \bibnamefont {Dou}}, \bibinfo {author} {\bibfnamefont {Xinru}\ \bibnamefont
  {Li}}, \bibinfo {author} {\bibfnamefont {Ying}\ \bibnamefont {Dai}}, \bibinfo
  {author} {\bibfnamefont {Baibiao}\ \bibnamefont {Huang}}, \ and\ \bibinfo
  {author} {\bibfnamefont {Yandong}\ \bibnamefont {Ma}},\ }\bibfield  {title}
  {\enquote {\bibinfo {title} {Strain-driven skyrmion--bimeron switching in
  topological magnetic monolayer crsebr},}\ }\href
  {https://doi.org/10.1039/D4MH00734D} {\bibfield  {journal} {\bibinfo
  {journal} {Materials Horizons}\ }\textbf {\bibinfo {volume} {11}},\ \bibinfo
  {pages} {5374--5380} (\bibinfo {year} {2024})}\BibitemShut {NoStop}%
\bibitem [{\citenamefont {Yu}\ \emph {et~al.}(2010)\citenamefont {Yu},
  \citenamefont {Onose}, \citenamefont {Kanazawa}, \citenamefont {Park},
  \citenamefont {Han}, \citenamefont {Matsui}, \citenamefont {Nagaosa},\ and\
  \citenamefont {Tokura}}]{yu2010real}%
  \BibitemOpen
  \bibfield  {author} {\bibinfo {author} {\bibfnamefont {XZ}~\bibnamefont
  {Yu}}, \bibinfo {author} {\bibfnamefont {Yoshinori}\ \bibnamefont {Onose}},
  \bibinfo {author} {\bibfnamefont {Naoya}\ \bibnamefont {Kanazawa}}, \bibinfo
  {author} {\bibfnamefont {Joung~Hwan}\ \bibnamefont {Park}}, \bibinfo {author}
  {\bibfnamefont {JH}~\bibnamefont {Han}}, \bibinfo {author} {\bibfnamefont
  {Yoshio}\ \bibnamefont {Matsui}}, \bibinfo {author} {\bibfnamefont {Naoto}\
  \bibnamefont {Nagaosa}}, \ and\ \bibinfo {author} {\bibfnamefont {Yoshinori}\
  \bibnamefont {Tokura}},\ }\bibfield  {title} {\enquote {\bibinfo {title}
  {Real-space observation of a two-dimensional skyrmion crystal},}\ }\href
  {https://doi.org/10.1038/nature09124} {\bibfield  {journal} {\bibinfo
  {journal} {Nature}\ }\textbf {\bibinfo {volume} {465}},\ \bibinfo {pages}
  {901--904} (\bibinfo {year} {2010})}\BibitemShut {NoStop}%
\bibitem [{\citenamefont {Tranchida}\ \emph {et~al.}(2018)\citenamefont
  {Tranchida}, \citenamefont {Plimpton}, \citenamefont {Thibaudeau},\ and\
  \citenamefont {Thompson}}]{tranchida2018massively}%
  \BibitemOpen
  \bibfield  {author} {\bibinfo {author} {\bibfnamefont {Julien}\ \bibnamefont
  {Tranchida}}, \bibinfo {author} {\bibfnamefont {Steven~J}\ \bibnamefont
  {Plimpton}}, \bibinfo {author} {\bibfnamefont {Pascal}\ \bibnamefont
  {Thibaudeau}}, \ and\ \bibinfo {author} {\bibfnamefont {Aidan~P}\
  \bibnamefont {Thompson}},\ }\bibfield  {title} {\enquote {\bibinfo {title}
  {Massively parallel symplectic algorithm for coupled magnetic spin dynamics
  and molecular dynamics},}\ }\href {https://doi.org/10.1016/j.jcp.2018.06.042}
  {\bibfield  {journal} {\bibinfo  {journal} {Journal of Computational
  Physics}\ }\textbf {\bibinfo {volume} {372}},\ \bibinfo {pages} {406--425}
  (\bibinfo {year} {2018})}\BibitemShut {NoStop}%
\bibitem [{\citenamefont {Thompson}\ \emph {et~al.}(2022)\citenamefont
  {Thompson}, \citenamefont {Aktulga}, \citenamefont {Berger}, \citenamefont
  {Bolintineanu}, \citenamefont {Brown}, \citenamefont {Crozier}, \citenamefont
  {In't~Veld}, \citenamefont {Kohlmeyer}, \citenamefont {Moore}, \citenamefont
  {Nguyen} \emph {et~al.}}]{thompson2022lammps}%
  \BibitemOpen
  \bibfield  {author} {\bibinfo {author} {\bibfnamefont {Aidan~P}\ \bibnamefont
  {Thompson}}, \bibinfo {author} {\bibfnamefont {H~Metin}\ \bibnamefont
  {Aktulga}}, \bibinfo {author} {\bibfnamefont {Richard}\ \bibnamefont
  {Berger}}, \bibinfo {author} {\bibfnamefont {Dan~S}\ \bibnamefont
  {Bolintineanu}}, \bibinfo {author} {\bibfnamefont {W~Michael}\ \bibnamefont
  {Brown}}, \bibinfo {author} {\bibfnamefont {Paul~S}\ \bibnamefont {Crozier}},
  \bibinfo {author} {\bibfnamefont {Pieter~J}\ \bibnamefont {In't~Veld}},
  \bibinfo {author} {\bibfnamefont {Axel}\ \bibnamefont {Kohlmeyer}}, \bibinfo
  {author} {\bibfnamefont {Stan~G}\ \bibnamefont {Moore}}, \bibinfo {author}
  {\bibfnamefont {Trung~Dac}\ \bibnamefont {Nguyen}},  \emph {et~al.},\
  }\bibfield  {title} {\enquote {\bibinfo {title} {Lammps-a flexible simulation
  tool for particle-based materials modeling at the atomic, meso, and continuum
  scales},}\ }\href {https://doi.org/10.1016/j.cpc.2021.108171} {\bibfield
  {journal} {\bibinfo  {journal} {Computer Physics Communications}\ }\textbf
  {\bibinfo {volume} {271}},\ \bibinfo {pages} {108171} (\bibinfo {year}
  {2022})}\BibitemShut {NoStop}%
\bibitem [{\citenamefont {Pajda}\ \emph {et~al.}(2001)\citenamefont {Pajda},
  \citenamefont {Kudrnovsk{\`y}}, \citenamefont {Turek}, \citenamefont
  {Drchal},\ and\ \citenamefont {Bruno}}]{pajda2001ab}%
  \BibitemOpen
  \bibfield  {author} {\bibinfo {author} {\bibfnamefont {Marek}\ \bibnamefont
  {Pajda}}, \bibinfo {author} {\bibfnamefont {J}~\bibnamefont
  {Kudrnovsk{\`y}}}, \bibinfo {author} {\bibfnamefont {Ilja}\ \bibnamefont
  {Turek}}, \bibinfo {author} {\bibfnamefont {Vaclav}\ \bibnamefont {Drchal}},
  \ and\ \bibinfo {author} {\bibfnamefont {Patrick}\ \bibnamefont {Bruno}},\
  }\bibfield  {title} {\enquote {\bibinfo {title} {Ab initio calculations of
  exchange interactions, spin-wave stiffness constants, and curie temperatures
  of fe, co, and ni},}\ }\href {https://doi.org/10.1103/PhysRevB.64.174402}
  {\bibfield  {journal} {\bibinfo  {journal} {Physical Review B}\ }\textbf
  {\bibinfo {volume} {64}},\ \bibinfo {pages} {174402} (\bibinfo {year}
  {2001})}\BibitemShut {NoStop}%
\bibitem [{\citenamefont {Cardias}\ \emph {et~al.}(2017)\citenamefont
  {Cardias}, \citenamefont {Szilva}, \citenamefont {Bergman}, \citenamefont
  {Marco}, \citenamefont {Katsnelson}, \citenamefont {Lichtenstein},
  \citenamefont {Nordstr{\"o}m}, \citenamefont {Klautau}, \citenamefont
  {Eriksson},\ and\ \citenamefont {Kvashnin}}]{cardias2017bethe}%
  \BibitemOpen
  \bibfield  {author} {\bibinfo {author} {\bibfnamefont {R}~\bibnamefont
  {Cardias}}, \bibinfo {author} {\bibfnamefont {Attila}\ \bibnamefont
  {Szilva}}, \bibinfo {author} {\bibfnamefont {Anders}\ \bibnamefont
  {Bergman}}, \bibinfo {author} {\bibfnamefont {I~Di}\ \bibnamefont {Marco}},
  \bibinfo {author} {\bibfnamefont {MI}~\bibnamefont {Katsnelson}}, \bibinfo
  {author} {\bibfnamefont {AI}~\bibnamefont {Lichtenstein}}, \bibinfo {author}
  {\bibfnamefont {Lars}\ \bibnamefont {Nordstr{\"o}m}}, \bibinfo {author}
  {\bibfnamefont {AB}~\bibnamefont {Klautau}}, \bibinfo {author} {\bibfnamefont
  {Olle}\ \bibnamefont {Eriksson}}, \ and\ \bibinfo {author} {\bibfnamefont
  {Yaroslav~O}\ \bibnamefont {Kvashnin}},\ }\bibfield  {title} {\enquote
  {\bibinfo {title} {The bethe-slater curve revisited; new insights from
  electronic structure theory},}\ }\href
  {https://doi.org/10.1038/s41598-017-04427-9} {\bibfield  {journal} {\bibinfo
  {journal} {Scientific reports}\ }\textbf {\bibinfo {volume} {7}},\ \bibinfo
  {pages} {4058} (\bibinfo {year} {2017})}\BibitemShut {NoStop}%
\bibitem [{SM()}]{SM}%
  \BibitemOpen
  \href@noop {} {}\bibinfo {note} {See Supplemental Material at [URL will be
  inserted by publisher] for details of the calculations of the exchange
  coupling under uniaxial strain, the effective couplings under small strain
  and supplementary figures and videos}\BibitemShut {NoStop}%
\bibitem [{\citenamefont {Shen}\ \emph {et~al.}(2022)\citenamefont {Shen},
  \citenamefont {Song}, \citenamefont {Xue}, \citenamefont {Wu}, \citenamefont
  {Wang},\ and\ \citenamefont {Zhong}}]{shen2022strain}%
  \BibitemOpen
  \bibfield  {author} {\bibinfo {author} {\bibfnamefont {Zhong}\ \bibnamefont
  {Shen}}, \bibinfo {author} {\bibfnamefont {Changsheng}\ \bibnamefont {Song}},
  \bibinfo {author} {\bibfnamefont {Yufei}\ \bibnamefont {Xue}}, \bibinfo
  {author} {\bibfnamefont {Zebin}\ \bibnamefont {Wu}}, \bibinfo {author}
  {\bibfnamefont {Jiqing}\ \bibnamefont {Wang}}, \ and\ \bibinfo {author}
  {\bibfnamefont {Zhicheng}\ \bibnamefont {Zhong}},\ }\bibfield  {title}
  {\enquote {\bibinfo {title} {Strain-tunable dzyaloshinskii-moriya interaction
  and skyrmions in two-dimensional janus cr2x3y3 (x, y= cl, br, i, x$\ne$ y)
  trihalide monolayers},}\ }\href {https://doi.org/10.1103/PhysRevB.106.094403}
  {\bibfield  {journal} {\bibinfo  {journal} {Physical Review B}\ }\textbf
  {\bibinfo {volume} {106}},\ \bibinfo {pages} {094403} (\bibinfo {year}
  {2022})}\BibitemShut {NoStop}%
\bibitem [{\citenamefont {Jiang}\ \emph {et~al.}(2022)\citenamefont {Jiang},
  \citenamefont {Huang}, \citenamefont {Zhu}, \citenamefont {Pan},
  \citenamefont {Fan}, \citenamefont {Zhang}, \citenamefont {Ma}, \citenamefont
  {Shi},\ and\ \citenamefont {Zhang}}]{jiang2022tuning}%
  \BibitemOpen
  \bibfield  {author} {\bibinfo {author} {\bibfnamefont {Lingzi}\ \bibnamefont
  {Jiang}}, \bibinfo {author} {\bibfnamefont {Can}\ \bibnamefont {Huang}},
  \bibinfo {author} {\bibfnamefont {Yan}\ \bibnamefont {Zhu}}, \bibinfo
  {author} {\bibfnamefont {Yanfei}\ \bibnamefont {Pan}}, \bibinfo {author}
  {\bibfnamefont {Jiyu}\ \bibnamefont {Fan}}, \bibinfo {author} {\bibfnamefont
  {Kaicheng}\ \bibnamefont {Zhang}}, \bibinfo {author} {\bibfnamefont
  {Chunlan}\ \bibnamefont {Ma}}, \bibinfo {author} {\bibfnamefont {Daning}\
  \bibnamefont {Shi}}, \ and\ \bibinfo {author} {\bibfnamefont {Hongbin}\
  \bibnamefont {Zhang}},\ }\bibfield  {title} {\enquote {\bibinfo {title}
  {Tuning the size of skyrmion by strain at the co/pt3 interfaces},}\ }\href
  {https://doi.org/10.1016/j.isci.2022.104039} {\bibfield  {journal} {\bibinfo
  {journal} {Iscience}\ }\textbf {\bibinfo {volume} {25}} (\bibinfo {year}
  {2022})}\BibitemShut {NoStop}%
\bibitem [{\citenamefont {Zhang}\ \emph
  {et~al.}(2025{\natexlab{a}})\citenamefont {Zhang}, \citenamefont {Xue},
  \citenamefont {Zhang}, \citenamefont {Hu}, \citenamefont {Wu},\ and\
  \citenamefont {Song}}]{zhang2025formation}%
  \BibitemOpen
  \bibfield  {author} {\bibinfo {author} {\bibfnamefont {Zhichao}\ \bibnamefont
  {Zhang}}, \bibinfo {author} {\bibfnamefont {Yufei}\ \bibnamefont {Xue}},
  \bibinfo {author} {\bibfnamefont {Qiuyao}\ \bibnamefont {Zhang}}, \bibinfo
  {author} {\bibfnamefont {Hongliang}\ \bibnamefont {Hu}}, \bibinfo {author}
  {\bibfnamefont {Xiaoping}\ \bibnamefont {Wu}}, \ and\ \bibinfo {author}
  {\bibfnamefont {Changsheng}\ \bibnamefont {Song}},\ }\bibfield  {title}
  {\enquote {\bibinfo {title} {The formation of elliptical n{\'e}el-type
  skyrmions via anisotropic exchange and dzyaloshinskii--moriya
  interactions},}\ }\href {https://doi.org/10.1063/5.0288764} {\bibfield
  {journal} {\bibinfo  {journal} {Applied Physics Letters}\ }\textbf {\bibinfo
  {volume} {127}} (\bibinfo {year} {2025}{\natexlab{a}})}\BibitemShut {NoStop}%
\bibitem [{\citenamefont {Paes}\ \emph {et~al.}(2019)\citenamefont {Paes},
  \citenamefont {Varalda},\ and\ \citenamefont {Mosca}}]{paes2019strain}%
  \BibitemOpen
  \bibfield  {author} {\bibinfo {author} {\bibfnamefont {VZC}\ \bibnamefont
  {Paes}}, \bibinfo {author} {\bibfnamefont {J}~\bibnamefont {Varalda}}, \ and\
  \bibinfo {author} {\bibfnamefont {DH}~\bibnamefont {Mosca}},\ }\bibfield
  {title} {\enquote {\bibinfo {title} {Strain-induced magnetization changes and
  magneto-volume effects in ferromagnets with cubic symmetry},}\ }\href
  {https://doi.org/10.1016/j.jmmm.2018.11.102} {\bibfield  {journal} {\bibinfo
  {journal} {Journal of Magnetism and Magnetic Materials}\ }\textbf {\bibinfo
  {volume} {475}},\ \bibinfo {pages} {539--543} (\bibinfo {year}
  {2019})}\BibitemShut {NoStop}%
\bibitem [{\citenamefont {Dos~Santos}\ \emph {et~al.}(2023)\citenamefont
  {Dos~Santos}, \citenamefont {Rom{\'a}}, \citenamefont {Tranchida},
  \citenamefont {Castedo}, \citenamefont {Cugliandolo},\ and\ \citenamefont
  {Bringa}}]{dossantos2023hysteresis}%
  \BibitemOpen
  \bibfield  {author} {\bibinfo {author} {\bibfnamefont {G}~\bibnamefont
  {Dos~Santos}}, \bibinfo {author} {\bibfnamefont {F}~\bibnamefont {Rom{\'a}}},
  \bibinfo {author} {\bibfnamefont {J}~\bibnamefont {Tranchida}}, \bibinfo
  {author} {\bibfnamefont {S}~\bibnamefont {Castedo}}, \bibinfo {author}
  {\bibfnamefont {LF}~\bibnamefont {Cugliandolo}}, \ and\ \bibinfo {author}
  {\bibfnamefont {Eduardo~Marcial}\ \bibnamefont {Bringa}},\ }\bibfield
  {title} {\enquote {\bibinfo {title} {Feasibility analysis towards the
  simulation of hysteresis with spin-lattice dynamics},}\ }\href
  {https://doi.org/10.1103/PhysRevB.108.134417} {\bibfield  {journal} {\bibinfo
   {journal} {Physical Review B}\ }\textbf {\bibinfo {volume} {108}},\ \bibinfo
  {pages} {134417} (\bibinfo {year} {2023})}\BibitemShut {NoStop}%
\bibitem [{\citenamefont {Zhang}\ \emph
  {et~al.}(2025{\natexlab{b}})\citenamefont {Zhang}, \citenamefont {Wu},
  \citenamefont {Zhang},\ and\ \citenamefont {Hu}}]{zhang2025uniaxial}%
  \BibitemOpen
  \bibfield  {author} {\bibinfo {author} {\bibfnamefont {Shuai}\ \bibnamefont
  {Zhang}}, \bibinfo {author} {\bibfnamefont {Yuanyuan}\ \bibnamefont {Wu}},
  \bibinfo {author} {\bibfnamefont {Zhao}\ \bibnamefont {Zhang}}, \ and\
  \bibinfo {author} {\bibfnamefont {Chengchao}\ \bibnamefont {Hu}},\ }\bibfield
   {title} {\enquote {\bibinfo {title} {Uniaxial strain modulation of
  multi-state skyrmion in fe3gate2},}\ }\href
  {https://doi.org/10.1142/S2010135X25400168} {\bibfield  {journal} {\bibinfo
  {journal} {Journal of Advanced Dielectrics}\ ,\ \bibinfo {pages} {2540016}}
  (\bibinfo {year} {2025}{\natexlab{b}})}\BibitemShut {NoStop}%
\bibitem [{\citenamefont {Essafi}\ \emph {et~al.}(2017)\citenamefont {Essafi},
  \citenamefont {Benton},\ and\ \citenamefont {Jaubert}}]{essafi2017}%
  \BibitemOpen
  \bibfield  {author} {\bibinfo {author} {\bibfnamefont {Karim}\ \bibnamefont
  {Essafi}}, \bibinfo {author} {\bibfnamefont {Owen}\ \bibnamefont {Benton}}, \
  and\ \bibinfo {author} {\bibfnamefont {L.~D.~C.}\ \bibnamefont {Jaubert}},\
  }\bibfield  {title} {\enquote {\bibinfo {title} {Generic nearest-neighbor
  kagome model: Xyz and dzyaloshinskii-moriya couplings with comparison to the
  pyrochlore-lattice case},}\ }\href {\doibase 10.1103/PhysRevB.96.205126}
  {\bibfield  {journal} {\bibinfo  {journal} {Phys. Rev. B}\ }\textbf {\bibinfo
  {volume} {96}},\ \bibinfo {pages} {205126} (\bibinfo {year}
  {2017})}\BibitemShut {NoStop}%
\bibitem [{\citenamefont {Ding}\ \emph {et~al.}(2025)\citenamefont {Ding},
  \citenamefont {Wang}, \citenamefont {Meng}, \citenamefont {Wan},
  \citenamefont {Wang}, \citenamefont {Xu}, \citenamefont {Zhu}, \citenamefont
  {Qin}, \citenamefont {Gao}, \citenamefont {Zhong} \emph
  {et~al.}}]{ding2025multistep}%
  \BibitemOpen
  \bibfield  {author} {\bibinfo {author} {\bibfnamefont {Bei}\ \bibnamefont
  {Ding}}, \bibinfo {author} {\bibfnamefont {Yadong}\ \bibnamefont {Wang}},
  \bibinfo {author} {\bibfnamefont {Jiahui}\ \bibnamefont {Meng}}, \bibinfo
  {author} {\bibfnamefont {Xuejin}\ \bibnamefont {Wan}}, \bibinfo {author}
  {\bibfnamefont {Qingping}\ \bibnamefont {Wang}}, \bibinfo {author}
  {\bibfnamefont {Xinxing}\ \bibnamefont {Xu}}, \bibinfo {author}
  {\bibfnamefont {Yu}~\bibnamefont {Zhu}}, \bibinfo {author} {\bibfnamefont
  {Minghui}\ \bibnamefont {Qin}}, \bibinfo {author} {\bibfnamefont {Xingsen}\
  \bibnamefont {Gao}}, \bibinfo {author} {\bibfnamefont {Xiaoyan}\ \bibnamefont
  {Zhong}},  \emph {et~al.},\ }\bibfield  {title} {\enquote {\bibinfo {title}
  {Multistep skyrmion phase transition driven by light-induced uniaxial
  strain},}\ }\href {https://doi.org/10.1126/sciadv.adt2698} {\bibfield
  {journal} {\bibinfo  {journal} {Science Advances}\ }\textbf {\bibinfo
  {volume} {11}},\ \bibinfo {pages} {eadt2698} (\bibinfo {year}
  {2025})}\BibitemShut {NoStop}%
\bibitem [{\citenamefont {Raab}\ \emph {et~al.}(2024)\citenamefont {Raab},
  \citenamefont {Schmitt}, \citenamefont {Brems}, \citenamefont {Roth\"orl},
  \citenamefont {Kammerbauer}, \citenamefont {Krishnia}, \citenamefont
  {Kl\"aui},\ and\ \citenamefont {Virnau}}]{Raab2024}%
  \BibitemOpen
  \bibfield  {author} {\bibinfo {author} {\bibfnamefont {Klaus}\ \bibnamefont
  {Raab}}, \bibinfo {author} {\bibfnamefont {Maurice}\ \bibnamefont {Schmitt}},
  \bibinfo {author} {\bibfnamefont {Maarten~A.}\ \bibnamefont {Brems}},
  \bibinfo {author} {\bibfnamefont {Jan}\ \bibnamefont {Roth\"orl}}, \bibinfo
  {author} {\bibfnamefont {Fabian}\ \bibnamefont {Kammerbauer}}, \bibinfo
  {author} {\bibfnamefont {Sachin}\ \bibnamefont {Krishnia}}, \bibinfo {author}
  {\bibfnamefont {Mathias}\ \bibnamefont {Kl\"aui}}, \ and\ \bibinfo {author}
  {\bibfnamefont {Peter}\ \bibnamefont {Virnau}},\ }\bibfield  {title}
  {\enquote {\bibinfo {title} {Skyrmion flow in periodically modulated
  channels},}\ }\href {\doibase 10.1103/PhysRevE.110.L042601} {\bibfield
  {journal} {\bibinfo  {journal} {Phys. Rev. E}\ }\textbf {\bibinfo {volume}
  {110}},\ \bibinfo {pages} {L042601} (\bibinfo {year} {2024})}\BibitemShut
  {NoStop}%
\bibitem [{\citenamefont {Zhang}\ \emph {et~al.}(2023)\citenamefont {Zhang},
  \citenamefont {Xia}, \citenamefont {Tretiakov}, \citenamefont {Ezawa},
  \citenamefont {Zhao}, \citenamefont {Zhou}, \citenamefont {Liu},\ and\
  \citenamefont {Mochizuki}}]{Zhang2023}%
  \BibitemOpen
  \bibfield  {author} {\bibinfo {author} {\bibfnamefont {Xichao}\ \bibnamefont
  {Zhang}}, \bibinfo {author} {\bibfnamefont {Jing}\ \bibnamefont {Xia}},
  \bibinfo {author} {\bibfnamefont {Oleg~A.}\ \bibnamefont {Tretiakov}},
  \bibinfo {author} {\bibfnamefont {Motohiko}\ \bibnamefont {Ezawa}}, \bibinfo
  {author} {\bibfnamefont {Guoping}\ \bibnamefont {Zhao}}, \bibinfo {author}
  {\bibfnamefont {Yan}\ \bibnamefont {Zhou}}, \bibinfo {author} {\bibfnamefont
  {Xiaoxi}\ \bibnamefont {Liu}}, \ and\ \bibinfo {author} {\bibfnamefont
  {Masahito}\ \bibnamefont {Mochizuki}},\ }\bibfield  {title} {\enquote
  {\bibinfo {title} {Laminar and transiently disordered dynamics of
  magnetic-skyrmion pipe flow},}\ }\href {\doibase 10.1103/PhysRevB.108.144428}
  {\bibfield  {journal} {\bibinfo  {journal} {Phys. Rev. B}\ }\textbf {\bibinfo
  {volume} {108}},\ \bibinfo {pages} {144428} (\bibinfo {year}
  {2023})}\BibitemShut {NoStop}%
\bibitem [{\citenamefont {Zhang}\ \emph
  {et~al.}(2025{\natexlab{c}})\citenamefont {Zhang}, \citenamefont {Xia},
  \citenamefont {Zhou}, \citenamefont {Zhao}, \citenamefont {Liu},
  \citenamefont {Xu},\ and\ \citenamefont {Mochizuki}}]{Zhang2025}%
  \BibitemOpen
  \bibfield  {author} {\bibinfo {author} {\bibfnamefont {Xichao}\ \bibnamefont
  {Zhang}}, \bibinfo {author} {\bibfnamefont {Jing}\ \bibnamefont {Xia}},
  \bibinfo {author} {\bibfnamefont {Yan}\ \bibnamefont {Zhou}}, \bibinfo
  {author} {\bibfnamefont {Guoping}\ \bibnamefont {Zhao}}, \bibinfo {author}
  {\bibfnamefont {Xiaoxi}\ \bibnamefont {Liu}}, \bibinfo {author}
  {\bibfnamefont {Yongbing}\ \bibnamefont {Xu}}, \ and\ \bibinfo {author}
  {\bibfnamefont {Masahito}\ \bibnamefont {Mochizuki}},\ }\bibfield  {title}
  {\enquote {\bibinfo {title} {Nanofluidic logic based on chiral skyrmion
  flows},}\ }\href {\doibase 10.1073/pnas.2506204122} {\bibfield  {journal}
  {\bibinfo  {journal} {Proceedings of the National Academy of Sciences}\
  }\textbf {\bibinfo {volume} {122}},\ \bibinfo {pages} {e2506204122} (\bibinfo
  {year} {2025}{\natexlab{c}})}\BibitemShut {NoStop}%
\bibitem [{\citenamefont {Kong}\ \emph {et~al.}(2023)\citenamefont {Kong},
  \citenamefont {Kov{\'a}cs}, \citenamefont {Charilaou}, \citenamefont {Zheng},
  \citenamefont {Wang}, \citenamefont {Han},\ and\ \citenamefont
  {Dunin-Borkowski}}]{kong2023direct}%
  \BibitemOpen
  \bibfield  {author} {\bibinfo {author} {\bibfnamefont {Deli}\ \bibnamefont
  {Kong}}, \bibinfo {author} {\bibfnamefont {Andr{\'a}s}\ \bibnamefont
  {Kov{\'a}cs}}, \bibinfo {author} {\bibfnamefont {Michalis}\ \bibnamefont
  {Charilaou}}, \bibinfo {author} {\bibfnamefont {Fengshan}\ \bibnamefont
  {Zheng}}, \bibinfo {author} {\bibfnamefont {Lihua}\ \bibnamefont {Wang}},
  \bibinfo {author} {\bibfnamefont {Xiaodong}\ \bibnamefont {Han}}, \ and\
  \bibinfo {author} {\bibfnamefont {Rafal~E}\ \bibnamefont {Dunin-Borkowski}},\
  }\bibfield  {title} {\enquote {\bibinfo {title} {Direct observation of
  tensile-strain-induced nanoscale magnetic hardening},}\ }\href
  {https://doi.org/10.1038/s41467-023-39650-8} {\bibfield  {journal} {\bibinfo
  {journal} {Nature Communications}\ }\textbf {\bibinfo {volume} {14}},\
  \bibinfo {pages} {3963} (\bibinfo {year} {2023})}\BibitemShut {NoStop}%
\end{thebibliography}

%

\end{document}